%% file: main.tex
\documentclass[letterpaper,onecolumn,unpublished]{quantumarticle}
\pdfoutput=1
\usepackage[most]{tcolorbox}
\usepackage{parskip} 
\usepackage{graphicx} 
\usepackage{siunitx}
\usepackage{pgfplots}
\usepackage{pdflscape}
\usepackage{listings}
\usepackage[T1]{fontenc}
\usepackage{braket}
\usepackage{booktabs}
\pgfplotsset{compat=1.18}
\usepackage{amsmath,amssymb,amsfonts,amsthm}
\usepackage[ruled,vlined]{algorithm2e}
\usepackage[numbers,sort&compress]{natbib}
\usepackage{xurl}
\usepackage[hidelinks]{hyperref}
\usepackage[capitalise,noabbrev]{cleveref}
\usepackage{nameref}

\newcommand{\F}{\mathbb{F}}
\newcommand{\C}{\mathbb{C}}
\newcommand{\Z}{\mathbb{Z}}
\newcommand{\legendre}[2]{\bigl(\tfrac{#1}{#2}\bigr)}

\newcommand{\hexval}[1]{\texttt{\color{violet!70!black}#1}}
\newcommand{\btc}{\begingroup\def\svgwidth{1.8ex}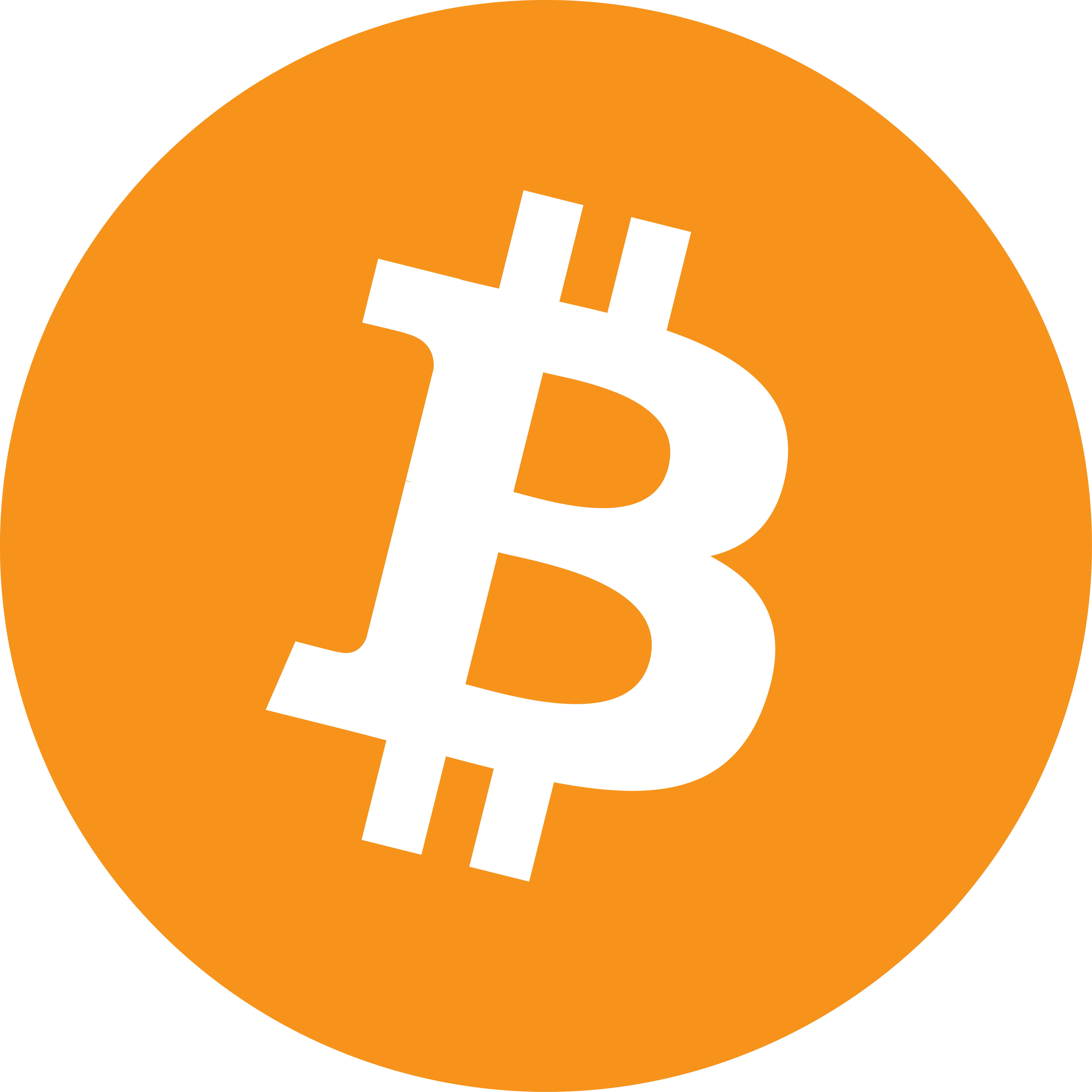\endgroup}
\newcommand{\secpk}[1]{\texttt{secp#1k1}}
\newcommand{\ECCDLcard}[5]{%
  \def\ifBTC{}%
  \ifnum#1=256\relax\def\ifBTC{\hspace{0.25em}\begingroup\def\svgwidth{1.8ex}\input{bitcoin-logo.pdf_tex}\endgroup}\fi
  \begin{tcolorbox}[colback=black!3,colframe=black!70,boxrule=0pt,arc=3pt,
    title=#1-bit \texttt{secp#1k1}\ifBTC,
    fonttitle=\bfseries\small]
  \ttfamily\scriptsize
  $p$:\\#2\\[0.25ex]
  $n$:\\#5\\[0.25ex]
  $P$:\\\hexval{#3}\\[0.25ex]
  $Q$:\\\hexval{#4}\\[0.25ex]
  $\kappa:\ P=[\kappa]Q$
  \end{tcolorbox}%
}

\definecolor{ibmBlue}  {RGB}{  0,114,178}   
\definecolor{gooTeal}  {RGB}{  0,158,115}   
\definecolor{qtmBrown} {RGB}{230,159,  0}   
\definecolor{ionMag}   {RGB}{204,121,167}   
\definecolor{naViolet} {RGB}{213, 94,  0}   
\definecolor{algGray}  {RGB}{ 90, 90, 90}   
\definecolor{myGold}   {RGB}{240,228, 66}   
\definecolor{darpaGr}  {RGB}{130,130,130}   
\definecolor{bsiGold}  {RGB}{204,121,  0}
\definecolor{eccBlack} {RGB}{  0,  0,  0}   

\definecolor{googleRed}{RGB}{213, 94,  0}   
\definecolor{quantGreen}{RGB}{  0,153,  0}   
\definecolor{neutralTeal}{RGB}{  0,158,115}   

\definecolor{darpaGold}{RGB}{204,121,  0}    
\definecolor{diraqCyan}{RGB}{  0,191,255}    

\newcommand{\hwach}[3]{%
  \addplot+[color=#1, solid, mark=#2, mark size=2.8pt, line width=1pt,
            mark options={draw=#1,fill=#1}, forget plot] coordinates{#3};}
\newcommand{\hwroad}[3]{%
  \addplot+[color=#1, dashed, mark=#2, mark size=2.8pt, line width=1pt,
            mark options={draw=#1,fill=white}, forget plot] coordinates{#3};}
\newcommand{\hwlegend}[2]{\addlegendimage{color=#1, solid, mark=#2, mark size=2.8pt, line width=1pt, mark options={draw=#1,fill=#1}}}

\newtheorem*{theorem*}{Theorem}
\newtheorem*{lemma*}{Lemma}
\newtheorem*{invariant*}{Invariant}

\theoremstyle{definition}

\theoremstyle{remark}

\begin{document}

\title{Brace for impact: ECDLP challenges for quantum cryptanalysis}
\author{Pierre-Luc Dallaire-Demers}
\orcid{0000-0002-9316-5597}
\email{pierre-luc@pauli.group}
\affiliation{Pauli Group}
\author{William Doyle}
\affiliation{Pauli Group}
\author{Timothy Foo}
\affiliation{Pauli Group}
\maketitle

\begin{abstract}
Precise suites of benchmarks are required to assess the progress of early fault-tolerant quantum computers at economically impactful applications such as cryptanalysis.
Appropriate challenges exist for factoring but those for elliptic curve cryptography are either too sparse or inadequate for standard applications of Shor's algorithm.
We introduce a difficulty-graded suite of elliptic curve discrete logarithm (ECDLP) challenges that use Bitcoin's curve $y^{2}=x^{3}+7 \pmod p$ while incrementally lowering the prime field from 256 down to 6 bits.
For each bit-length, we provide the prime, the prime group order, and two deterministic nothing-up-my-sleeve (NUMS) points in compressed SEC1 form.
All challenges are generated by a deterministic, reproducible procedure, and no private challenge scalar is chosen in advance.
We calibrate classical cost against Pollard's rho records and quantum cost against resource estimation results for Shor's algorithm.
We compile Shor's ECDLP circuit to logical counts and map them to physical resources for various parameters of the surface code, the repetition cat code and the LDPC cat codes. 
Under explicit and testable assumptions on physical error rates, code distances, and non-Clifford supply, our scenarios place the full 256-bit instance within a 2027--2033 window.
The challenge ladder thus offers a transparent ruler to track fault-tolerant progress on a cryptanalytic target of immediate relevance, and it motivates proactive migration of digital assets to post-quantum signatures.
\end{abstract}
\clearpage

\hfill
\begin{figure}[ht]
\centering
\begin{tikzpicture}
\begin{axis}[
    width=0.6\linewidth,
    height=0.45\linewidth,
    axis lines=middle,
    xlabel={$x$},
    ylabel={$y$},
    xmin=-5,xmax=5,
    ymin=-5,ymax=5,
    domain=-1.912931183:3.0,
    samples=1000,
    axis equal image,
    tick label style={font=\small},
    xticklabel=\empty,
    yticklabel=\empty,
]
  \addplot[thick] {sqrt(x^3 + 7)};
  \addplot[thick] {-sqrt(x^3 + 7)};
  \addplot[thick] coordinates {(-1.912931183,-0.28) (-1.912931183,0.28)};
\end{axis}
\end{tikzpicture}
\caption{Real locus of $y^{2}=x^{3}+7$. The plot is for geometric intuition only and does not reflect arithmetic over $\F_{p}$.}
\label{fig:real-mordell}
\end{figure}

\hfill
\tableofcontents
\clearpage

\section{Introduction \label{sec:Introduction}}
With error-corrected ``megaquop'' machines, quantum computers capable of roughly one million logical operations, now plausibly within reach, it is timely to map the landscape of first commercial-scale applications for fault-tolerant quantum computing (FTQC) \cite{Preskill2025_Megaquop}.
Among these, Shor's algorithm for elliptic-curve discrete logarithms (ECDLP) is especially noteworthy because it combines immediate economic relevance with comparatively modest FTQC requirements.
Because Bitcoin relies on ECDSA over \texttt{secp256k1}, sufficiently capable FTQC threatens private keys once public keys appear on chain \cite{Nakamoto2008,SEC2v2,Aggarwal2018}.
The prudent response is to prepare migration paths to post-quantum signatures.

\begin{figure}[t]
\centering
\includegraphics[width=0.92\linewidth]{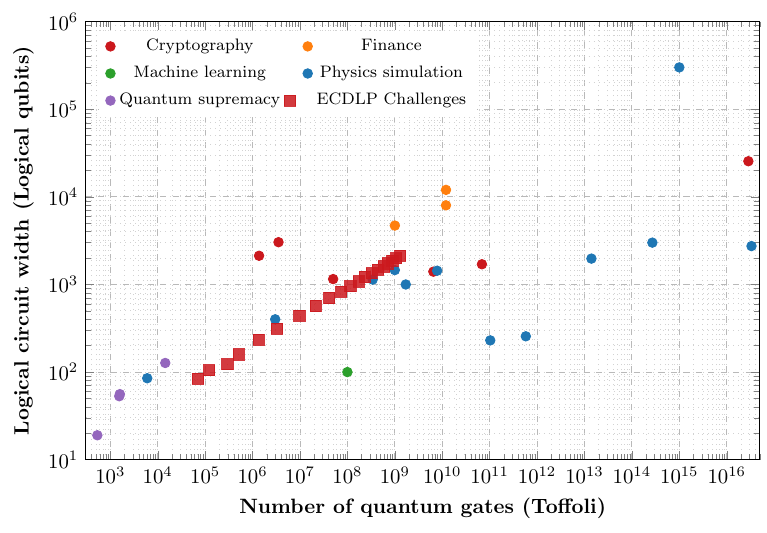}
\caption{\textbf{Representative FTQC applications in logical-width/Toffoli-count space.}
Filled markers denote achieved demonstrations; open markers denote published resource estimates.
The plot is intended as an orienting map rather than a complete survey.}
\label{fig:applicationsLandscape}
\end{figure}

Figure~\ref{fig:applicationsLandscape} situates representative FTQC workloads from quantum-supremacy and utility experiments, finance, machine learning, physics and chemistry simulation, and cryptanalysis on a common logical-width/Toffoli-count plane \cite{Arute2019_Sycamore53,Wu2021_StrongAdvantage,Kim2023_UtilityBeforeFT,DerivativePricing2021_Q,FinancialRisk2022_Q,DerivativePricingQSP2024_Q,TDA2024_PRXQuantum,CytochromeP4502023_PNAS,Beverland2022,OpenSystemSimulation2024,FermiHubbard2024,NMRSpectralPrediction2024,QREChem2024,TensorFactorization2025,SampleBasedKrylov2025,SpectrumAmplification2025,Haner2020,Chevignard2024,Litinski2023_ECC,garn2025quantum,Gidney2025_RSA2048}.
On this broader map, ECDLP lies near the low end of the resource spectrum among economically relevant FTQC targets, which helps explain why it is a plausible early cryptanalytic application to monitor.

What signals can the community use to track real progress toward that threshold?
We need simple, public, reproducible rulers.
Existing benchmarks do not provide sufficient resolution. The Certicom ECC Challenges remain historically important but too sparse to register year-over-year advances \cite{Certicom2009}.
Fine-grained “Bitcoin puzzles” and Proof-of-Quantum restrict the secret to an interval and therefore map to Pollard's kangaroo rather than to Shor's period finding over the full 256-bit group order \cite{PonsZieniewicz2020,BTCPuzzle_List_2025,Doyle2025_ProofOfQuantum}.
A size-graded suite of ECDLP challenges can serve as canaries: imperfect, because disclosures depend on what teams publish and because FTQC progress can arrive abruptly, but still the best early warning we can maintain and audit.
No group has yet run Shor's ECDLP end-to-end on a prime field, but hardware trajectories suggest first demonstrations may be close.

We therefore keep Bitcoin's curve
\begin{equation}
E/\F_{p}: y^{2}=x^{3}+7 \pmod p
\label{eqn:secp256k1}
\end{equation}
and its semantics, and only shrink the prime field. The curve is plotted over the real numbers in Fig.\ref{fig:real-mordell}.
For each of twenty target field bit-lengths between 6 and 256 we provide $(p,P_p,Q_p)$ with prime group order, a deterministic recipe to regenerate the parameters, and reference logical resource estimates for Shor's ECDLP circuit.
This ladder, from a 6-bit toy to the full 256-bit instance, offers a transparent, fine-grained ruler for both classical and fault-tolerant quantum progress and a concrete impetus to migrate digital assets to post-quantum signatures.

The contributions are as follows.
(i) A deterministic, reproducible recipe for a ladder of $\texttt{secp256k1}$-shaped ECDLP instances ranging from 6 to 256 bits, together with code to regenerate all parameters.
(ii) A calibrated classical baseline that cleanly separates generic ECDLP from interval-restricted puzzles and anchors claims to published records.
(iii) A logical-level cost model for Shor's ECDLP circuit tied to concrete reversible arithmetic choices, and a physical mapping under repetition and LDPC cat codes as well as surface codes.
(iv) A sensitivity analysis that makes plain which physical and architectural levers move the 256-bit point across the 2027-2033 window.
(v) A public ruler for reporting and comparing future progress on quantum cryptanalysis of elliptic curves.

To orient the reader, we first motivate \texttt{secp256k1} as a natural benchmark (Section~\ref{sub:BitcoinECDSA}), then recall the relevant elliptic-curve background and the ECDLP (Sections~\ref{sub:EllipticCurveCryptography}--\ref{sub:PollardRho}).
We next summarize the minimal quantum computing tools (Section~\ref{sub:QuantumComputing}) and instantiate Shor's algorithm for ECDLP (Section~\ref{sub:ShorECDLP}).
With those pieces in hand we introduce a ladder of challenge instances (Section~\ref{sec:Challenges}), calibrate classical cost (Section~\ref{sec:ClassicalResources}), and map quantum resource requirements and roadmaps (Section~\ref{sec:QuantumResources}).
We close with a discussion of implications, limitations, and open directions (Section~\ref{sec:Discussion}).

\subsection{Bitcoin's \texttt{secp256k1} as an FTQC benchmark}
\label{sub:BitcoinECDSA}

Bitcoin uses digital signatures to authorize the spending of coins.
In practice, the reference implementation employs ECDSA over the \texttt{secp256k1} curve, whose domain parameters are specified by the Standards for Efficient Cryptography Group (SECG) and whose curve equation is given by \eqref{eqn:secp256k1} with a well-known base point and prime group order.
\footnote{We use \texttt{secp256k1} throughout as a special-form short-Weierstrass curve with $A=0,B=7$; parameters per \cite{SEC2v2}. Background on special-form curves appears in \cite{SafeCurves2013}.}
At a protocol level, signatures bind transactions to keys; at an algorithmic level, they reduce to scalar multiplication and verification on a prime-order subgroup of $E(\F_{p})$ \cite{Nakamoto2008,SEC2v2}.

The “Patoshi'' pattern of early Bitcoin mining is believed to be associated with Satoshi Nakamoto, the pseudonymous creator of Bitcoin, contains at least 21,953 blocks with 1,097,650 BTC in total \cite{Sigman2025_PatoshiAddresses}.

Because \texttt{secp256k1} is both widely deployed and precisely specified, it is a natural benchmark for early fault-tolerant quantum computing (FTQC).
It fixes a target problem (ECDLP at 256 bits) with unambiguous correctness criteria, rich test vectors, and an immediate path to cross-check classical versus quantum resource estimates.
The DARPA Quantum Benchmarking program pursues the same aim: place algorithms and hardware on a common ruler so we can read progress at a glance \cite{DARPA_QBenchmark2021}.
The risk motivation is equally clear.
If sufficiently large fault-tolerant machines become available, Shor's algorithm threatens ECDSA and therefore the security of Bitcoin addresses exposed on chain \cite{Aggarwal2018}.
Beyond cryptanalytic feasibility, the crypto-economic consequences of a credible quantum capability have been explored in market-design terms \cite{Rohde2021_QCryptoEcon}, reinforcing the value of a transparent, standardized benchmark like \texttt{secp256k1} for engineering coordination.

\subsection{Elliptic-curve cryptography and the ECDLP}
\label{sub:EllipticCurveCryptography}
Having fixed the protocol context and why \texttt{secp256k1} serves as a natural ruler, we now recall the minimal elliptic-curve background and define the ECDLP precisely.
This fixes notation and isolates the group-theoretic assumptions that drive the classical and quantum analyses that follow.
Over a prime field $\F_{p}$, a (short-Weierstrass) elliptic curve is the set of points $(x,y)\in\F_{p}^{2}$ satisfying $y^{2}=x^{3}+Ax+B$ together with a point at infinity, provided the discriminant $\Delta=-16(4A^{3}+27B^{2})\not\equiv0\pmod p$.
This set forms an abelian group under a geometric chord-and-tangent law; in software we work with explicit rational formulas.
A point $G$ of large prime order 
\begin{equation}
n := \lvert E(\F_{p})\rvert
\label{eqn:group-order}
\end{equation}
generates a cyclic subgroup $\langle G\rangle\subseteq E(\F_{p})$.
The number of bits to represent the group order is
\begin{equation}
b := \left\lceil \log_{2} n \right\rceil .
\label{eqn:bitlength}
\end{equation}

Scalar multiplication $[k]G$ means adding $G$ to itself $k$ times \cite{Galbraith2012}.
It can be used to define a public key
\begin{equation}
Q=[d]G
\label{eqn:pubkey}
\end{equation}
as used in ECDSA.

The security of such cryptography rests on the elliptic curve discrete logarithm problem (ECDLP): given $G$ and $Q$, recover the scalar $d$.
There is no known classical sub-exponential algorithm for generic groups of this form.
The best general-purpose classical algorithm is Pollard's rho, which needs on the order of $O(\sqrt{n})$ group operations and uses only modest memory.
This is roughly $2^{b/2}$ steps \cite{Galbraith2012}.
Hence the 256-bit case is far beyond classical reach, while smaller instances are useful for calibration.

Practical elliptic-curve cryptography (ECC) standardization requires that the group order $n$ has a large prime factor to avoid Pohlig-Hellman reductions.
In our challenges we follow the usual discipline: choose $p$ so that the curve has (or contains) a prime-order subgroup and then work inside that subgroup \cite{Koblitz1988,Zywina2011,Certicom2009}.
This aligns directly with ECDSA's security assumptions and keeps comparisons with classical and quantum algorithms clean.
With these definitions in hand, we turn next to the best generic classical attack, Pollard's rho (Section~\ref{sub:PollardRho}), to establish a concrete baseline.

\subsection{Pollard's rho for ECDLP}
\label{sub:PollardRho}

With the group structure fixed, a natural question is how hard discrete logarithms remain on classical machines.
Pollard's rho sets the benchmark and also explains the record computations we cite later.

Pollard's rho for ECDLP performs a pseudo-random walk over group elements represented as
\begin{equation}
X_i = [a_i]G + [b_i]Q .
\label{eqn:pollard-walk}
\end{equation}
A simple partition function (e.g., by a few low bits of $X_i$) selects one of several update rules that algebraically update $(a_i,b_i)$ and $X_i$ at each step.
When two distinct indices $i\neq j$ produce a collision $X_i=X_j$, we obtain $[a_i-a_j]G=[b_j-b_i]Q$, which reveals $d$ by a single modular inversion in $\Z_{n}$ provided $b_j\neq b_i$ \cite{Galbraith2012}:
\begin{equation}
[a_i-a_j]G = [b_j-b_i]Q \quad \Longrightarrow \quad
d \equiv (a_i-a_j)\,(b_j-b_i)^{-1} \pmod n .
\label{eqn:pollard-solve}
\end{equation}

GPUs, made abundant by AI workloads, reduce time-per-iteration and increase attainable parallelism; in the \texttt{secp256k1} interval setting this has moved the state of the art from 114 bits in 2020 to 129 bits by 2025 \cite{BTCPuzzle_List_2025}.

The expected number of group operations is
\begin{equation}
\mathbb{E}[T_{\rho}] \approx \sqrt{\frac{\pi n}{2}},\qquad
\mathbb{E}[T_{\rho}^{\pm}] \approx \sqrt{\frac{\pi n}{4}},\qquad
T_{\rho}=\Theta(2^{b/2}) .
\label{eqn:rho-expected}
\end{equation}
with the second expression obtained when the negation map is exploited in prime-order subgroups. Memory remains modest with cycle-finding; massively parallel implementations rely on distinguished points.

\begin{algorithm}[H]
\DontPrintSemicolon\small
\SetAlgoLined
Choose partition function $S:E(\F_p)\to\{1,\dots,m\}$ and update rules $U_j(X)=[\alpha_j]X+[ \beta_j ]G+[ \gamma_j ]Q$.\;
Pick random $a_0,b_0\in\Z_n$, set $X_0=[a_0]G+[b_0]Q$.\;
\For{$i=0,1,2,\dots$}{
  $j\leftarrow S(X_i)$;\quad $X_{i+1}\leftarrow U_j(X_i)$;\quad update $(a_{i+1},b_{i+1})$ accordingly.\;
  if distinguished$(X_{i+1})$ then report $X_{i+1},a_{i+1},b_{i+1}$.\;
}
Upon collision $X_i=X_j$ with $i\neq j$, output $d\equiv (a_i-a_j)(b_j-b_i)^{-1}\pmod n$ if $b_j\neq b_i$.
\caption{Pollard's rho method for ECDLP.}
\label{alg:PollarRho}
\end{algorithm}

Expected runtime is $\Theta(\sqrt{\pi n/2})$ group operations, matching the birthday bound up to a constant factor.
Memory can be kept low with cycle-finding \cite{Floyd1967,Brent1980}, or reduced to communication overhead using “distinguished points'' in massively parallel settings.
These techniques have powered several record computations and provide an empirical ruler for classical progress \cite{Galbraith2012,BernsteinEtAl2016}.

Variants such as Pollard's kangaroo target keys known to lie in an interval, which is relevant for “puzzle'' addresses or constrained keyspaces.
Well-engineered implementations on consumer hardware and field-programmable gate arrays (FPGAs) illustrate the practical scaling: a 112-bit prime-field curve on game consoles \cite{Bos2012}, a 113-bit Koblitz curve on an FPGA cluster \cite{Wenger2014}, and a 114-bit interval on \texttt{secp256k1} with the kangaroo method \cite{PonsZieniewicz2020}.
More recently, puzzles up to 129 bits have been solved \cite{BTCPuzzle_List_2025}; while these puzzles can be tackled with Pollard's rho, they are not appropriate for Shor's algorithm since the underlying group is still the full 256-bit curve.
These points anchor the classical side of Fig.~\ref{fig:classicalECDLP}.
Keys restricted to a range of size $2^{t}$ admit Pollard's kangaroo in $O(2^{t/2})$ steps; Shor's algorithm for ECDLP does not benefit from such range restrictions and still requires period finding modulo the full group order $n$.

Two decades of engineering effort have not displaced the generic $\Theta(2^{b/2})$ barrier for prime-field curves; practical wins come from constant-factor improvements and parallelism rather than new asymptotics.
In the 2015 curve-standardization debate, Koblitz and Menezes emphasized the need for transparency about assumptions and long-horizon risks, including the prospect of large-scale quantum computers, helping frame why it is natural to place classical baselines alongside quantum resource estimates \cite{KoblitzMenezes2015_Riddle}.
With this motivation, we now review the basics of quantum computing.

\subsection{Basics of quantum computing}
\label{sub:QuantumComputing}

The classical picture above sets the bar for how hard the problem is.
The next two subsections provide the minimal quantum toolkit and then the specific instantiation of Shor's algorithm for ECDLP.

A qubit is a two-level quantum system whose state is a vector in $\C^{2}$.
Computation proceeds by unitary gates (reversible, norm-preserving linear maps) interleaved with measurements in specified bases.
Entanglement and interference give rise to computational behaviors without classical analogues; the canonical references for formalism and models of computation are \cite{NielsenChuang2010}.

Two algorithmic speedups are most relevant for cryptography.
Grover's algorithm quadratically accelerates unstructured search, impacting brute-force key search but not breaking public-key schemes outright.
Shor's algorithm gives an \emph{exponential} speedup for factoring and discrete logarithms in abelian groups, threatening Rivest-Shamir-Adleman (RSA) and ECC in principle.
Realizing these asymptotics at scale requires fault tolerance: encoding logical qubits into many physical qubits and executing long circuits with active error correction.

Surface code architectures with lattice surgery are the most widely studied path to large-scale FTQC.
Resource estimates map algorithmic cost (logical width/depth, $T$-count) into physical-qubit counts and wall-time via a choice of code distance and non-Clifford supply (distillation or cultivation) \cite{Litinski2019}.
We adopt this standard costing model when comparing classical and quantum paths to ECDLP across bit-lengths.

In the next subsection we specialize these ingredients to Shor's algorithm for ECDLP and identify the logical building blocks needed for resource estimates (Section~\ref{sub:ShorECDLP}). 
Beyond public-key schemes, Grover's algorithm halves the effective security of symmetric primitives; prudent “hybrid'' defenses double key sizes (e.g., AES-256, SHA-384/512) and prefer PQ KEMs/signatures in protocols.
For ECDLP in generic prime-order groups, no quantum algorithm asymptotically improves on Shor; quantum walks yield limited constant-factor effects only for structured families, which our instances avoid \cite{Galbraith2012}.
We therefore treat Shor as the relevant asymptotic baseline, and Grover as a parallel concern for protocol designers.

\subsection{Shor's algorithm for ECDLP}
\label{sub:ShorECDLP}

Equipped with a brief quantum-computing primer, we now specialize Shor's hidden-subgroup machinery to elliptic curves.

At heart, Shor's algorithm reduces discrete logarithms to period finding in a finite abelian group \cite{Shor1994}.
For ECDLP we fix a generator $G$ of order $n$ and a public key $Q$.
Prepare two control registers and a point register, and implement the unitary
\begin{equation}
U_{G,Q}:\ \ket{x}\ket{y}\ket{P}\ \mapsto\ \ket{x}\ket{y}\ket{P+[x]G+[y]Q} .
\label{eqn:U-GQ}
\end{equation}
Measuring the point register yields some coset value, which collapses the control pair $(x,y)$ to a uniform superposition constrained by $x+d\,y\equiv \text{const}\pmod n$.

Applying the quantum Fourier transform (QFT) modulo $n$ to the two control registers reveals samples $(x',y')$ that satisfy 
\begin{equation}
y' \equiv d\,x' \pmod n .
\label{eqn:shor-sample}
\end{equation}
in the ideal exact-$\mathrm{QFT}_n$ circuit.
For prime $n$, any sample with $x'\neq 0$ yields $d=y'(x')^{-1}\bmod n$; the zero sample occurs with probability $1/n$ and is discarded.
The success probability and sample complexity follow the usual hidden-subgroup analysis \cite{Proos2003}.

A convenient instantiation is as follows: 
Prepare two control registers of $b=\lceil\log_2 n\rceil$ qubits each, with combined control width
\begin{equation}
n_{e} = 2\bigl\lceil \log_{2} n \bigr\rceil .
\label{eqn:control-size}
\end{equation}
and a point register $\ket{\mathcal{O}}$.
Starting from uniform coherent superpositions over $x,y\in\{0,\dots,n-1\}$, compute
$U_{G,Q}\ket{x}\ket{y}\ket{\mathcal{O}}=\ket{x}\ket{y}\ket{[x]G+[y]Q}$, measure the point register, apply $\mathrm{QFT}_{n}$ to each control, and measure to obtain $(x',y')$ satisfying $y'\equiv d\,x' \pmod n$ with high probability.
A nonzero $x'$ determines $d$ by modular inversion, followed by the check $[d]G=Q$.
Implementations using power-of-two Fourier transforms instead require the corresponding approximate-sampling postprocessing \cite{Proos2003}.

\begin{algorithm}[H]
\DontPrintSemicolon\small
\SetAlgoLined
\KwIn{$E/\F_p: y^{2}=x^{3}+7$; generator $G\in E(\F_p)$ of prime order $n$; public key $Q=[d]G$.}
\KwOut{$d\in\{1,\dots,n-1\}$ such that $Q=[d]G$.}
$b \leftarrow \lceil \log_2 n \rceil$;\quad $n_e\leftarrow 2b$ \tcp*[f]{combined control width}\;
\While{true}{
  \textbf{Prepare registers:} $X$ of $b$ qubits; $Y$ of $b$ qubits; point register $P$.\;
  \textbf{Initialize:} $P\leftarrow\ket{\mathcal O}$; 
  $X \leftarrow \frac{1}{\sqrt n}\sum_{x=0}^{n-1}\ket{x}$;\ 
  $Y \leftarrow \frac{1}{\sqrt n}\sum_{y=0}^{n-1}\ket{y}$.\;
  \textbf{Compute:} apply $U_{G,Q}$ so that 
  \(\ket{x}\ket{y}\ket{P}\mapsto \ket{x}\ket{y}\ket{P+[x]G+[y]Q}\).\;
  \textbf{Measure} $P$ and discard outcome. 
  \tcp*[f]{$(X,Y)$ collapses to a random coset of $x+d\,y\equiv c\pmod n$}\;
  \textbf{Fourier step:} apply $\mathrm{QFT}_n$ to $X$ and to $Y$.\;
  \textbf{Measure} $X,Y$ to obtain $(u,v)\in\{0,\dots,n-1\}^{2}$, with $v\equiv d\,u\pmod n$.\;
  \If(\tcp*[f]{classical post-proc.}){$u\not\equiv 0\pmod n$}{
     $d^\star \leftarrow v\,u^{-1}\bmod n$ \tcp*[f]{one modular inverse over $\Z_n$}\;
     \If{\([d^\star]G = Q\)}{\Return $d^\star$}
  }
}
\caption{Ideal exact-$\mathrm{QFT}_n$ version of Shor's algorithm for the elliptic curve discrete logarithm over $\langle G\rangle\subseteq E(\F_p)$.}
\label{alg:ShorECDLP}
\end{algorithm}

In practice, the dominant engineering task is implementing $U_{G,Q}$ reversibly and fault-tolerantly: reversible field arithmetic, point addition/doubling circuits, and careful qubit accounting. 
Concrete logical-level resource counts and subsequent optimizations appear in \cite{Roetteler2017_ECC,Haner2020}; these feed directly into the physical-level costing summarized in Fig.~\ref{fig:resourcesECDLP} later in the paper.

These resource drivers motivate the size-graded challenge suite we introduce next (Section~\ref{sec:Challenges}).
\clearpage
\section{Creating the challenges \label{sec:Challenges}}

Having set the algorithmic target, we now turn it into a concrete ladder of instances that preserve the \texttt{secp256k1} curve form while varying the field size.
This ruler lets us connect algorithmic cost to both classical records and hardware trajectories.

We present a ladder of elliptic-curve discrete-logarithm challenges that keeps the curve shape used by Bitcoin but shrinks the prime field.
The curve is always given by \eqref{eqn:secp256k1}, and only the field prime $p$ varies with the target bit-length.
For a published bit-length $k\in\mathcal K$, where
\[
\mathcal K=\{6,8,12,16,24,32\}\cup\{48,64,80,\ldots,256\},
\]
we denote the corresponding instance by \secpk{k} (e.g., \secpk{32} for $k=32$).
The field bit length $k=\lceil\log_2 p\rceil$ and group-order bit length $b=\lceil\log_2 n\rceil$ are distinct, even though the cofactor is 1.
For example, the 6-bit card has $n=31$ and $b=5$, whereas the 64-bit card has $b=65$.
Except at 256 bits, these labels are shorthand for our custom prime-field ladder rather than the standardized SEC2 curves of the same names.


Each challenge package contains four items: a $k$-bit prime $p$; the prime group order $n=|E(\F_{p})|$; and two deterministically derived NUMS points $P_{p},Q_{p}\in E(\F_{p})$.
The embedding degree is defined as the multiplicative order of $p$ modulo $n$,
\begin{equation}
r := \operatorname{ord}_{n}(p)\ \text{ such that }\ p^{r}\equiv 1 \pmod n .
\label{eqn:embed-degree}
\end{equation}

Bitcoin itself uses ECDSA over the standardized curve \texttt{secp256k1}, specified by SECG \cite{SEC2v2}.
It fixes a short-Weierstrass equation \eqref{eqn:secp256k1} over a 256-bit prime field together with a base point $G$ of large prime order and comprehensive test vectors.
Our challenges keep this exact curve shape but scale $p$ down so that classical experiments and fault-tolerant resource studies can be performed across a wide range of sizes without changing the underlying group law.

Rather than sampling a secret scalar, we derive the published points from the fixed ASCII labels \texttt{"Quantum"} and \texttt{"Challenge"}.
For any label $L$, define
\begin{align}
h_L &= \operatorname{SHA256}(L) \bmod p, \nonumber\\
\delta_L &= \min\Bigl\{\, d\ge 0:\ \legendre{(h_L+d)^{3}+7}{p}\in\{0,1\}\Bigr\}, \nonumber\\
x_L &\equiv h_L+\delta_L \pmod p,\qquad
y_L^{2}\equiv x_L^{3}+7\pmod p,\qquad
R_L=(x_L,y_L).
\label{eqn:nums-point}
\end{align}
We choose the canonical root $y_L$ in the range $0\le y_L\le (p-1)/2$.
The SEC1 compressed encoding of $R_L$ then uses prefix \texttt{02} when $y_L$ is even and \texttt{03} when $y_L$ is odd.
We set $P_p$ to the point derived from \texttt{"Quantum"} and $Q_p$ to the point derived from \texttt{"Challenge"}.
Because $n=|E(\F_p)|$ is prime, every non-identity point generates $E(\F_p)$; thus $Q_p$ is a generator and there exists a unique $\kappa\in\{1,\dots,n-1\}$ such that $P_p=[\kappa]Q_p$.
This deterministic construction eliminates any hidden or sampled private key while leaving the ECDLP hardness unchanged.
For Shor's algorithm we simply take the generator to be $Q_p$ and the target point to be $P_p$.
All published points below are written in compressed SEC1 form for compactness.

\noindent
For clarity, the deterministic construction is summarized below.

\begin{algorithm}[H]
\DontPrintSemicolon\small
\SetAlgoLined
\textbf{Input}: a published field bit-length $k\in\mathcal K$; curve $E/\F_{p}: y^{2}=x^{3}+7$; labels \texttt{"Quantum"} and \texttt{"Challenge"}.\\
\textbf{Output}: $p$ (a $k$-bit prime), $n=|E(\F_{p})|$ (prime), $r=\mathrm{ord}_{n}(p)$, and NUMS points $P_p,Q_p\in E(\F_p)$ such that $P_p=[\kappa]Q_p$ for a unique $\kappa\in\{1,\dots,n-1\}$.\\[0.25ex]
\textbf{Prime search.}\;
\eIf{$k=256$}{
  $p\leftarrow 2^{256}-2^{32}-977$ \tcp*[f]{standard secp256k1 field}\;
  $n\leftarrow |E(\F_p)|$ (SEA or equivalent)\;
}{
  Enumerate $k$-bit primes in ascending order for $k\in\{6,8,12,16\}$, and descending order otherwise\;
  \For{each candidate $p$ in this order}{
    compute $n\leftarrow |E(\F_{p})|$ for $E:y^{2}=x^{3}+7$ (SEA or equivalent)\;
    \If{$n$ is prime $\wedge\ n\neq p$}{accept this $p$; \textbf{break}.}
  }
}
\textbf{Embedding degree.}\;
$r\leftarrow \min\{\,t\ge 1:\ p^{t}\equiv 1 \pmod n\,\}$.\;
\textbf{NUMS points.}\;
\ForEach{$(L,R)\in\{(\texttt{"Quantum"},P_p),(\texttt{"Challenge"},Q_p)\}$}{
  $h\leftarrow \operatorname{SHA256}(L)\bmod p$;\quad $\delta\leftarrow 0$.\;
  \While{true}{
    $x\leftarrow h+\delta \bmod p$;\quad $\rho\leftarrow x^{3}+7 \bmod p$.\;
    \If{$\rho$ is a square in $\F_p$}{
      choose $y\in\F_p$ with $y^{2}\equiv \rho\pmod p$ and $0\le y\le (p-1)/2$\;
      $R\leftarrow(x,y)$;\quad \textbf{break}\;
    }
    $\delta\leftarrow \delta+1$\;
  }
}
\textbf{Return} $p,n,r,P_p,Q_p$.
\caption{Deterministic construction of the \secpk{k} challenge instance.}
\label{alg:ChallengeConstruction}
\end{algorithm}

The full list of challenges is provided in Sec.\ref{sec:ECDLPchallenges} of the Appendix.
In each card we list $p$, $P_p$, $Q_p$, and the prime group order $n$.
The 6-bit case is the most trivial instance and can be solved by hand by a motivated reader. Yet, its resolution in a complete fault-tolerant implementation of Shor's algorithm will be a milestone.
A natural mid-scale rung is the 160-bit instance.
The most economically impactful instance is the 256-bit instance, which uses the actual \texttt{secp256k1} field prime together with our deterministic NUMS point pair.

These instances form the common ruler for the remainder of the paper.
In Section~\ref{sec:ClassicalResources} we calibrate classical cost via Pollard's rho and the public record of ECDLP computations.
In Section~\ref{sec:QuantumResources} we map Shor's algorithm onto logical and physical resources.

\section{Challenges calibration: Classical resources \label{sec:ClassicalResources}}
Before translating Shor's algorithm into qubits, we first calibrate the ruler against the best public classical computations.
This sets lower bounds for any purported advantage and anchors the scale of our plots.

Classical progress on the elliptic-curve discrete logarithm problem (ECDLP) has been measured against a small set of public milestones.
The Certicom ECC Challenge provided the common ruler: carefully curated problem instances with clear correctness criteria and rewards, which catalyzed comparable algorithms and implementations across platforms \cite{Certicom2009}.
A key early waypoint was the community solution of the binary curve ECC2-109: Monico and collaborators demonstrated a massively distributed Pollard-rho with distinguished points and effective use of the negation map, establishing the basic engineering template for large ECDLP computations \cite{Monico2004}.
The subsequent break of ECC2K-130 by Bailey \emph{et al.} and the ECC Challenge team extended these techniques to Koblitz curves over binary fields, exploiting Frobenius endomorphisms and careful load balancing across heterogeneous hardware \cite{Bailey2009,EccChallenge2011}.
On prime-field curves, Bos \emph{et al.} solved a 112-bit instance using thousands of PlayStation~3 consoles; the work popularized high-throughput Pollard-rho with “sloppy” modular reduction and efficient single-precision arithmetic to cut constant factors \cite{Bos2012}.
Wenger and Wolfger then showed that FPGAs are an excellent match to the Pollard update function: a 113-bit Koblitz-curve discrete log fell to a tightly pipelined, multi-FPGA cluster with low-overhead distinguished-point reporting \cite{Wenger2014}.
Their follow-up refined the hardware datapath and memory system, highlighting how bit-serial/bit-sliced arithmetic and careful partitioning of the walk reduce time-per-iteration on reconfigurable logic \cite{Wenger2016}.
In parallel, Bernstein \emph{et al.} engineered faster FPGA designs and host-side software for ECDLP, quantifying how far constant-factor improvements and better parallel orchestration can push classical records without changing asymptotics \cite{BernsteinEtAl2016}.
Finally, Pons and Zieniewicz used Pollard's kangaroo to crack a 114-bit \emph{interval} on \texttt{secp256k1}; this is not a full 256-bit discrete log, but it cleanly illustrates the best-in-class performance one can expect when the key is known to lie in a prescribed range \cite{PonsZieniewicz2020}.

\begin{figure}[ht]
    \centering
    \begin{tikzpicture}
        \begin{axis}[
            width=\linewidth,
            height=0.6\textheight,
            xlabel={Year},
            ylabel={Curve size (bits)},
            width=10cm,height=9cm,
            xmin=1996,xmax=2025,
            xtick={1996,2000,2004,2008,2012,2016,2020,2024},
            xticklabels={1996,2000,2004,2008,2012,2016,2020,2024},
            ymin=76,ymax=130,ytick={76,80,...,130},
            grid=both,
            tick label style={font=\small},
            only marks,
            mark=*,
            mark options={fill=blue},
            nodes near coords,
            point meta=explicit symbolic,
            visualization depends on={value \thisrow{dx} \as \dx},
            visualization depends on={value \thisrow{dy} \as \dy},
            every node near coord/.style={anchor=center,font=\scriptsize,
                                            xshift=\dx pt,yshift=\dy pt},
        ]

        \addplot table[
        col sep=comma,
        x=year, y=bits, meta=lbl
        ] {resource_estimator/classical_ecdlp_records.dat};

        \end{axis}
    \end{tikzpicture}
    \caption{Classical ECDLP challenges. Prime-field records cluster near 112--113 bits; interval-restricted \texttt{secp256k1} results have advanced to 129 bits on GPU clusters.}
    \label{fig:classicalECDLP}
\end{figure}

\begin{figure}[ht]
    \centering
    \begin{tikzpicture}
        \begin{axis}[
        xlabel={Bit-strength $b$ (bits)},
        ylabel={Pollard's rho operations},
        xmin=0,xmax=256, xtick distance=32,
        ymode=log,log basis y=10,
        ymin=1e0,ymax=1e39,
        ytick={1e0,1e6,1e12,1e18,1e24,1e30,1e36},
        yticklabel style={/pgf/number format/sci},
        grid=both,
        grid style={gray!50},
        extra y ticks={
            1e8,            
            6e9,            
            3.6e11,         
            8.64e12,        
            2.592e14,       
            3.15576e15,     
            3.15576e18,     
            3.15576e21,     
            3.15576e24      
        },
        extra y tick labels={
            1\,s,
            1\,min,
            1\,h,
            1\,d,
            1\,mo,
            1\,yr,
            1\,kyr,
            1\,Myr,
            1\,Gyr
        },
        extra y tick style={
            xshift=1.5cm,grid style={dotted,gray,xshift=-1.5cm},major tick length=0pt
        },
        ]
        \addplot[thick, mark=*, color=black, mark options={fill=blue,draw=blue}] coordinates {
            (6,    8)
            (8,    1.6e1)
            (12,   64)
            (16,   2.56e2)
            (24,   4.096e3)
            (32,   6.554e4)
            (48,   1.678e7)
            (64,   4.295e9)
            (80,   1.100e12)
            (96,   2.815e14)
            (112,  7.206e16)
            (128,  1.845e19)
            (144,  4.722e21)
            (160,  1.209e24)
            (176,  3.095e26)
            (192,  7.923e28)
            (208,  2.028e31)
            (224,  5.192e33)
            (240,  1.329e36)
            (256,  3.403e38)
        };
        \end{axis}
    \end{tikzpicture}
    \caption{Number of classical operations and approximate runtime under a fixed single-core normalization for the ECDLP challenges using Pollard's rho; large GPU deployments shift the curve downward roughly in proportion to aggregate throughput but do not alter the $\Theta(2^{b/2})$ slope.}    
    \label{fig:classicalECDLPruntime}
\end{figure}

Read together, Figs.~\ref{fig:classicalECDLP} and~\ref{fig:classicalECDLPruntime} tell a consistent story.
The timeline shows a rapid climb from the late 1990s through the mid-2000s: 79, 89, 95, 97, then 108--109 bits, driven by better Pollard-rho engineering, the use of automorphisms (negation map, Frobenius on Koblitz curves), and large-scale parallelism.
After 2009 the pace slows: prime-field \emph{secp112r1} at 112 bits, a 113-bit Koblitz curve on FPGAs, a mid-2010s binary-field record in roughly the 117-bit range, and importantly, an \emph{interval} attack at 114 bits on \texttt{secp256k1}.
The qualitative plateau beyond $\sim\!120$ bits is visible in the sparse right-hand side of Fig.~\ref{fig:classicalECDLP}.

The runtime plot explains why.
For a group of order $\approx 2^{b}$, generic classical attacks scale as $\Theta(2^{b/2})$ group operations.
Moving from 112 to 120 bits multiplies the required work by roughly $2^{(120-112)/2}=16$; moving from 112 to 128 bits costs a factor of $2^{8}=256$.
The curve at 112 bits sits near $7.2\times10^{16}$ group operations in Fig.~\ref{fig:classicalECDLPruntime}; at a fixed single-core reference rate this corresponds to decades, which real-world records overcome only by enlisting vast numbers of devices.
Constant-factor advances, vectorization, “sloppy” reduction, endomorphism speedups on special curves, better partition functions, and distinguished-point infrastructures shift the curve downward, and near-linear parallelization via distinguished points reduces wall-clock time by throwing hardware at the problem.
But each additional $\sim\!8$ bits demands $\times 16$ more aggregate work, so hardware, energy, memory-bandwidth, and interconnect budgets quickly dominate; coordination overheads and I/O for collision reporting further erode ideal speedups at very large scales.

This is why the classical record exhibits a fast early regime and then flattens.
Early wins benefited from smaller instances, curve structure (binary and Koblitz curves admit cheap endomorphisms), and the easy gains from moving to FPGAs, graphics processing units (GPUs), and console clusters.
Once those constant factors were harvested, the $2^{b/2}$ barrier reasserted itself.  Interval-restricted results (kangaroo) and structured-curve records remain valuable calibration points, but they are not directly comparable to a generic 256-bit prime-field ECDLP.

These constraints on generic classical attacks motivate looking beyond brute force; the next section translates the ECDLP circuit into logical and physical quantum resources (Section~\ref{sec:QuantumResources}).

In light of these data, it is reasonable to treat $\approx\!120$ bits as an effective boundary for \emph{generic} classical ECDLP under realistic resource budgets: below this line, carefully engineered projects have succeeded; above it, the exponential wall implied by Fig.~\ref{fig:classicalECDLPruntime} requires aggregate effort that grows beyond practicality.


\section{Challenges calibration: Quantum resources \label{sec:QuantumResources}}

The goal of this section is to translate Shor's algorithm for the elliptic-curve discrete logarithm problem (ECDLP) into physically testable requirements.
The translation proceeds in two steps.
First, we specify the \emph{logical} circuit model: the number of logical qubits $N_{\mathrm{log}}$, the Toffoli count $T_{\mathrm{count}}$, and the Toffoli depth $T_{\mathrm{depth}}$ produced by reversible elliptic-curve arithmetic \cite{Roetteler2017_ECC,Haner2020}.
Second, we map these logical resources to a hardware footprint and a wall-clock duration under three error-correction settings: a surface-code baseline with lattice surgery \cite{Fowler2009_SurfaceCodeCosts,fowler2012surface,Litinski2019}, a repetition cat code architecture \cite{Gouzien2023_ECC}, and an LDPC cat code architecture \cite{ruiz2025ldpc}.
The mapping depends on a small set of experimentally exposed parameters: physical gate error rates, cycle time, connectivity, and non-Clifford resource supply.

\subsection{Logical resources}

Shor's ECDLP uses a period-finding scaffold specialized to the group $\langle G\rangle\subseteq E(\F_p)$ \cite{Shor1994,Proos2003}.
The logical workload is set by reversible field arithmetic in point addition and doubling.
We follow Häner-Jaques-Naehrig-Roetteler-Soeken (HJNRS) \cite{Haner2020}: windowed arithmetic with explicit space-time trade-offs yields a triplet that depends on both the field width $k$ and the group-order width $b$,
\[
N_{\mathrm{log}}(k,b),\qquad T_{\mathrm{count}}(k,b),\qquad T_{\mathrm{depth}}(k,b),
\]
where Toffolis are the non-Clifford bottleneck.
The existing resource grids use a single nominal size $k$ for both arithmetic and exponent lengths, hence the approximation $b\approx k$ and $2k$ exponent bits.
They are size-based estimates, not exact compilations of every Appendix~\ref{sec:ECDLPchallenges} card; exact-instance costing must use the actual pair $(k,b)$, which differs for eight cards.
A newer width-optimized alternative due to Chevignard, Fouque, and Schrottenloher compresses the ECDLP output via a Legendre-symbol hash and residue-number-system arithmetic; their detailed P-256 tables report $1193$ logical qubits, but at the cost of $22$ runs and about $2^{39.98}$ Toffolis per run \cite{Chevignard2026_ECDLP}.
Because our physical tables target practical runtime and factory demand rather than minimum width at any cost, we retain the HJNRS family as the baseline logical workload.
In our costing Toffolis are supplied by factories (distillation or cultivation) and all other Clifford operations are budgeted at logical cycle granularity \cite{Litinski2019,Gidney2024_MSCultivation}.
For bookkeeping we bound the overall failure probability of a run by
\begin{equation}
\varepsilon_{\mathrm{run}} \;\le\; T_{\mathrm{ops}}\, p_{L}(d) \;\le\; \varepsilon_{\mathrm{target}},
\label{eq:budget}
\end{equation}
where $T_{\mathrm{ops}}$ is the number of fault-tolerant opportunities for a logical fault (dominated by non-Clifford steps under our schedules), $p_{L}(d)$ is the per-op logical error at code distance $d$, and $\varepsilon_{\mathrm{target}}$ is a fixed engineering target (e.g. $10^{-2}$ to $10^{-3}$ total failure).

\subsection{Surface-code mapping}

For two-dimensional (2D) surface codes operated below threshold $p_{\mathrm{th}}$, the logical error rate decreases exponentially with distance $d$.
A standard approximation in the regime $p\ll p_{\mathrm{th}}$ is
\begin{equation}
p_{L}(d) \;\approx\; C\!\left(\frac{p}{p_{\mathrm{th}}}\right)^{\frac{d+1}{2}},
\label{eq:plogic}
\end{equation}
with $C=\mathcal{O}(10^{-1})$, $p$ a representative circuit-level physical error rate, and $p_{\mathrm{th}}\sim 10^{-2}$ for the surface code \cite{fowler2012surface}.
Equations \eqref{eq:budget}-\eqref{eq:plogic} set the smallest $d$ that meets the run failure target for a given hardware point.
With lattice surgery, the footprint separates into data patches and non-Clifford factories:
\begin{equation}
N_{\mathrm{phys}} \;\simeq\; \alpha\, d_{\mathrm{data}}^{\,2}\,N_{\mathrm{log}} \;+\; \beta\, d_{\mathrm{fac}}^{\,2}\, F,
\label{eq:nphys}
\end{equation}
where $\alpha,\beta$ are layout constants, $d_{\mathrm{data}}$ and $d_{\mathrm{fac}}$ are the distances chosen for data and factory patches, and $F$ is the number of parallel factories \cite{Litinski2019}.
The wall time is the maximum of a depth-limited term and a supply-limited term:
\begin{equation}
t \;=\; \max\!\Bigl\{\, c\, d_{\mathrm{data}}\,T_{\mathrm{depth}}\,\tau \;,\;\; \frac{T_{\mathrm{count}}}{F\, r_{\mathrm{fac}}(d_{\mathrm{fac}})}\,\tau \Bigr\},
\label{eq:runtime}
\end{equation}
with cycle time $\tau$, schedule constant $c$, and per-cycle Toffoli (or $T$-state) rate $r_{\mathrm{fac}}$ provided by the factories \cite{Litinski2019,GidneyEkera2019_RSA2048,Gidney2024_MSCultivation}.
For concreteness we report two stylized hardware points implemented in our in-house surface-code post-processing model. The conservative point takes $p=10^{-3}$, $\tau=1.5\,\mu\mathrm{s}$, $(\gamma_T,\gamma_C)=(12,8)$, $(\alpha_{\mathrm{data}},\alpha_{\mathrm{fac}})=(2.5,25)$, $k_{\mathrm{fac}}=80$, and $25\%$ time/qubit margins. The aggressive point takes $p=2\times 10^{-4}$, $\tau=0.2\,\mu\mathrm{s}$, $(\gamma_T,\gamma_C)=(6,4)$, $(\alpha_{\mathrm{data}},\alpha_{\mathrm{fac}})=(1.5,10)$, $k_{\mathrm{fac}}=25$, and $10\%$ margins. These choices reduce distances and shift both \eqref{eq:nphys} and \eqref{eq:runtime} downward in the aggressive case \cite{Litinski2019,Gidney2024_MSCultivation}.
The Microsoft/Q\# resource-count pipeline supplies the logical inputs, while the physical surface-code translation reported here is our own transparent post-processing script rather than a direct run of the modern Microsoft Resource Estimator \cite{Soeken2023_CalcResCryptanalysis}.

\subsection{Repetition cat code}

Cat qubits encode a logical qubit in superpositions of coherent states of a bosonic mode and can be engineered to exhibit strongly biased noise: bit flips are exponentially suppressed by two-photon dissipation while phase flips dominate \cite{Gouzien2023_ECC}.
A repetition cat code combats the dominant phase errors by a one-dimensional repetition of $d$ cats with majority voting.
Writing $p_{Z}$ for the per-cycle phase-flip probability on a single cat and assuming independent errors, the logical phase-flip probability after one cycle is
\begin{equation}
p^{(Z)}_{L,\mathrm{rep}}(d) \;=\; \sum_{j=\lceil(d+1)/2\rceil}^{d} \binom{d}{j}\, p_{Z}^{\,j}\,(1-p_{Z})^{d-j},
\label{eq:rep-cat-logical}
\end{equation}
while bit flips are exponentially rare by design. The encoding is one-dimensional and dispenses with the $d^{2}$ area law of the surface code, so the memory footprint scales linearly in $d$ for fixed logical accuracy.
Under experimentally motivated parameters (cycle time near 500\,ns, strong bias), a full 256-bit ECDLP instance has been costed at about $126{,}133$ cat qubits with a nine-hour runtime \cite{Gouzien2023_ECC}.
This places the repetition cat code well below surface-code baselines at comparable failure targets while keeping wall time in the same hours regime.
Compiling ECDLP to this architecture uses the same HJNRS logical circuits; the change is entirely in the physical layer: biased noise, phase-focused checks, and factories adapted to bias-preserving gadgets.

\subsection{LDPC cat code}

Quantum LDPC codes achieve higher encoding rates by using sparse, typically low-weight parity checks.
When combined with cat qubits, LDPC codes can be specialized to correct phase flips with strictly local checks in 2D layouts.
The LDPC cat code introduced by Ruiz, Guillaud, Leverrier, Mirrahimi, and Vuillot concatenates a cat layer with a classical LDPC code that protects the logical phase, preserving 2D locality and enabling significantly higher rate than the surface code \cite{ruiz2025ldpc}.
The net effect is a large reduction of memory footprint for a fixed run failure target.
Using the same logical workload and the same failure budgeting as above, and following the open methodology and code paths in \cite{Gouzien2024_elliptic_log_cat}, our estimator places the 256-bit ECDLP instance at 38581 cat qubits under the aggressive hardware point. Depending on decoder latency and factory scheduling, the wall time ranges from hours to about a day. Unfolded and bias-aware distillation schemes further reduce non-Clifford overheads in this setting \cite{Ruiz2025_UnfoldedDistillation}.

\subsection{Anchors at 6 bits and 256 bits}

At six bits the algorithm is small enough to serve as an end-to-end FTQC demonstration target.
The logical instance fits in well under a few hundred logical qubits with a modest Toffoli budget; Equation~\eqref{eq:plogic} then yields single-digit distances at conservative error rates.
Both repetition cat code and LDPC cat code run this instance in seconds on $\mathcal{O}(10^{3})$ or fewer cat qubits under the hardware points above, while an aggressive surface-code baseline fits below $10^{4}$ physical qubits with sub-minute runtime.
The six-bit rung is therefore ideal for the first complete demonstrations of Shor's ECDLP with active error correction.

At 256 bits the spread of estimates reflects architectural leverage.
Conservative surface-code assumptions push the footprint into the multi-million-qubit regime and multi-day runtime, consistent with RSA-scale estimates \cite{GidneyEkera2019_RSA2048}.
Aggressive surface-code assumptions move toward sub-million qubits and hours-scale runs if factory throughput is improved \cite{Litinski2019,Gidney2024_MSCultivation}.
The repetition cat code achieves about $1.26\times 10^{5}$ cat qubits and a nine-hour run under published parameters \cite{Gouzien2023_ECC}.
The LDPC cat code reduces the memory further, below $4\times 10^{4}$ cat qubits under aggressive but explicit assumptions, at the cost of tighter scheduling and decoder performance \cite{ruiz2025ldpc,Gouzien2024_elliptic_log_cat,Ruiz2025_UnfoldedDistillation}.
These brackets motivate the view that ECDLP is a candidate for the first wave of FTQC applications on “megaquop” machines \cite{Preskill2025_Megaquop}.

\subsection{Interpretation}

Equations \eqref{eq:budget}-\eqref{eq:runtime} isolate the levers that matter.
Improving the physical-to-threshold ratio $p/p_{\mathrm{th}}$ reduces $d$ and hence area quadratically; faster cycle time $\tau$ compresses wall time linearly; better non-Clifford throughput $F\,r_{\mathrm{fac}}$ eliminates supply bottlenecks.
Biased bosonic qubits change the geometry entirely by shifting protection to the dominant error channel and by enabling high-rate LDPC protection with strictly local checks.
Within this framework, the present, reasonable range for a 256-bit ECDLP break runs from millions of physical qubits (conservative surface code) down to a few times $10^{4}$ cat qubits (LDPC cat code), with repetition cat code in between \cite{Fowler2009_SurfaceCodeCosts,fowler2012surface,Litinski2019,GidneyEkera2019_RSA2048,Gouzien2023_ECC,ruiz2025ldpc}.
Width-optimized ECDLP variants can lower logical qubit counts further, but current constructions do so by increasing total non-Clifford work by orders of magnitude and therefore do not tighten this practical physical range \cite{Chevignard2026_ECDLP}.
The next section reviews the historical evolution of these estimates and plots crossing timelines under the same modeling assumptions.

\begin{figure}[ht]
    \centering
    \begin{minipage}[t]{0.48\linewidth}
        \centering
        \begin{tikzpicture}
          \begin{axis}[%
            xlabel={Logical qubits},
            ylabel={Toffoli gates},
            xmode=log,
            ymode=log,
            ymin=1e4, ymax=1e10,
            width=6cm, height=6cm,
            only marks,
            mark size=2.5pt,
            colormap name=viridis,
            point meta min=6, point meta max=256,
          ]
          \addplot+[
            scatter, only marks,
            mark=square*,
            scatter src=explicit,
          ] table [meta=id] {resource_estimator/logical_series_lowW.dat};
          \addplot+[
            scatter, only marks,
            mark=triangle*,
            scatter src=explicit,
          ] table [meta=id] {resource_estimator/logical_series_lowT.dat};
          \addplot+[
            scatter, only marks,
            mark=*,
            scatter src=explicit,
          ] table [meta=id] {resource_estimator/logical_series_lowD.dat};

          \addplot[
            nodes near coords,
            point meta=explicit symbolic,
            only marks, mark=none,
            every node near coord/.style={
              font=\scriptsize,
              fill=white, fill opacity=0.85, text opacity=1, inner sep=1pt
            },
          ] table [x=x, y=y, meta=lbl] {resource_estimator/logical_anchor_labels.dat};

          \end{axis}
        \end{tikzpicture}
    \end{minipage}
    \hfill
    \begin{minipage}[t]{0.48\linewidth}
        \centering
        \begin{tikzpicture}
          \begin{axis}[%
            xlabel={Physical qubits},
            ylabel={Time},
            xmode=log,
            ymode=log,
            ymin=1e-1, ymax=1e7,
            ytick={1,60,3600,86400,604800,2592000},
            yticklabels={1 s,1 min,1 hr,1 day,1 week,1 month},
            yticklabel pos=right,
            ylabel near ticks,
            width=6cm, height=6cm,
            only marks,
            mark size=2.5pt,
            colormap name=viridis,
            point meta min=6, point meta max=256,
            colorbar,
            colorbar style={
              title={bits $k$},
              ytick={6,32,64,96,128,160,192,224,256},
            },
          ]
          \addplot[
            scatter, only marks,
            mark=*,
            scatter src=explicit,
          ] table [meta=id] {resource_estimator/repetition_cat_series.dat};
          \addplot+[
            scatter, only marks,
            mark=square*,
            scatter src=explicit,
          ] table [meta=id] {resource_estimator/ldpc_cat_series.dat};

          \addplot[scatter, only marks, mark=triangle*, scatter src=explicit]
          table [meta=id] {resource_estimator/physical_series_lowT_aggressive_fixed.dat};
          \addplot[scatter, only marks, mark=triangle, scatter src=explicit]
          table [meta=id] {resource_estimator/physical_series_lowT_conservative_fixed.dat};
          \addplot[scatter, only marks, mark=diamond*, scatter src=explicit]
          table [meta=id] {resource_estimator/physical_series_lowW_aggressive_fixed.dat};
          \addplot[scatter, only marks, mark=diamond, scatter src=explicit]
          table [meta=id] {resource_estimator/physical_series_lowW_conservative_fixed.dat};
          \addplot[scatter, only marks, mark=star, scatter src=explicit]
          table [meta=id] {resource_estimator/physical_series_lowD_aggressive_fixed.dat};
          \addplot[scatter, only marks, mark=+, scatter src=explicit]
          table [meta=id] {resource_estimator/physical_series_lowD_conservative_fixed.dat};

          \addplot[
            nodes near coords,
            point meta=explicit symbolic,
            only marks, mark=none,
            every node near coord/.style={
              font=\scriptsize,
              fill=white, fill opacity=0.85, text opacity=1, inner sep=1pt
            },
          ] table [x=x, y=y, meta=lbl] {resource_estimator/physical_anchor_labels.dat};
          \draw[dotted, thin] ({axis cs:38581,0}|-{rel axis cs:0,0}) -- ({axis cs:38581,0}|-{rel axis cs:0,1});

          \end{axis}
        \begin{scope}[overlay]
        \end{scope}
        \end{tikzpicture}
    \end{minipage}

    \caption{\textbf{Logical and physical resources to break $k$-bit ECDLP with Shor's algorithm.}
    \emph{Left:} logical width vs.\ Toffoli count from \cite{Haner2020} following various optimization procedures. Low width (square), low T gates count (triangle), and low depth (circles).
    \emph{Right:} repetition-cat (circles) and LDPC-cat (squares) resource estimates from \cite{Gouzien2024_elliptic_log_cat}; triangles, diamonds, and stars show the conservative and aggressive surface-code baselines.
    Color encodes the bit‑size $k$.}
    \label{fig:resourcesECDLP}
\end{figure}

\subsection{Evolution of resource estimation \label{sub:EvolutionResources}}

With the costing model fixed, we can trace how algorithmic refinements and improved non-Clifford supply have steadily driven down qubit requirements.
Shor's introduction of polynomial-time quantum algorithms for factoring and discrete logarithms via period finding and the quantum Fourier transform established that RSA and DLP lie in BQP \cite{Shor1994}.
Proos and Zalka provided the first detailed adaptation of Shor's discrete logarithm algorithm to elliptic curves over prime fields, giving explicit reversible circuits and early resource estimates \cite{Proos2003}.
Gate-level accounting by Roetteler \emph{et al.} delivered widely used width, depth, and Toffoli/T baselines for ECDLP over prime fields \cite{Roetteler2017_ECC}.
Häner \emph{et al.} then improved the scalar multiplication kernel through windowing, adaptive uncomputation, and a faster modular inversion based on a reformulated binary Euclidean algorithm, cutting both T-count and T-depth \cite{Haner2020}.
Chevignard, Fouque, and Schrottenloher later adapted output compression and residue-number-system arithmetic directly to elliptic curves, roughly halving the width of the P-256 instance in their detailed tables, but only by paying a multi-run and high-gate-count penalty; the result is therefore best read as a width optimization, not yet a lower-cost practical attack \cite{Chevignard2026_ECDLP}.
Architectural translations mapped these logical schedules to concrete platforms: Webber \emph{et al.} adapted Gidney's throughput analysis to Shor's algorithm for ion-trap architectures \cite{Webber2022};
Gouzien \emph{et al.} produced a full-stack estimate on a repetition cat code architecture, finding that ECC-256 can be solved in $\sim$ 9 hours with $126{,}133$ cat qubits \cite{Gouzien2023_ECC};
and Litinski's active volume framework, together with batch inversion and state reuse, reduced the ECC-256 workload to $\sim 5\times10^{7}$ Toffoli gates and highlighted large speedups over strictly 2D local layouts \cite{Litinski2023_ECC}.

In parallel, end-to-end factoring studies established the canonical methodology for converting logical resources into surface code costs.
Gidney and Eker{\aa} integrated lattice surgery surface codes \cite{Litinski2019} with magic state factories to show that RSA-2048 could be factored in about eight hours using $\sim 2\times 10^{7}$ noisy qubits under realistic cycle times \cite{GidneyEkera2019_RSA2048}.
More recent work combines approximate residue arithmetic due to Chevignard-Fouque-Schrottenloher, yoked surface code storage, and, crucially, cheaper non-Clifford supply via magic state cultivation, bringing the RSA-2048 requirement below one million noisy qubits at multi-day runtimes under comparable physical assumptions \cite{Chevignard2024,Gidney2024_MSCultivation,Gidney2025_RSA2048}.
Webster \emph{et al.} then introduced the Pinnacle QLDPC architecture and estimated RSA-2048 factoring below $10^{5}$ physical qubits at $p=10^{-3}$, $1\,\mu$s code cycles, and $10\,\mu$s reaction time \cite{Webster2026_Pinnacle}.
These RSA results sharpen the architecture side of the warning signal, but absent a comparably detailed ECDLP compilation on Pinnacle they should not be read as a revised ECC-256 point estimate.

\begin{figure}[ht]
\centering
\begin{tikzpicture}
\begin{semilogyaxis}[
  xmin=2008,xmax=2035,
  xtick      ={2010,2015,2020,2025,2030,2035},
  xticklabels={2010,2015,2020,2025,2030,2035},
  xticklabel style={font=\small,rotate=45,anchor=east,yshift=-0.5ex},
  xlabel={\textbf{Year}},
  ylabel={\textbf{Physical qubits}},
  ymin=1,ymax=1e9,
  ytick={1,10,100,1e3,1e4,1e5,1e6,1e7,1e8,1e9},
  yticklabels={1,10,100,1\,k,10\,k,100\,k,1\,M,10\,M,100\,M,1\,B},
  log ticks with fixed point,
  grid=both,
  minor grid style={dotted,gray!30},
  major grid style={dashed,gray!50},
  every axis plot/.append style={forget plot},
  legend style={
    font=\footnotesize,
    at={(0.5,-0.38)},anchor=north,
    draw=none,
    column sep=6pt,row sep=2pt,
  },
  legend image post style={solid},
  legend columns=3,
  legend cell align=left,
  set layers=standard,
]
\begin{pgfonlayer}{axis background}
  \fill[gray!15] (axis cs:2027,1) rectangle (axis cs:2033,1e9);
\end{pgfonlayer}

\addplot+[only marks,mark=*,mark size=2.8pt,color=algGray,mark options={draw=algGray,fill=algGray},opacity=0.6]
  coordinates{(2012,1e9)(2019,2e7)(2023,349133)(2025,1e6)(2026,9.8e4)(2026,1.3255e4)};

\addplot+[only marks,mark=square*,mark size=2.8pt,color=eccBlack,
          mark options={draw=eccBlack,fill=eccBlack},opacity=0.9]
  coordinates{(2017,5.8e8)(2023,6.912e6)(2023,1.26133e5)(2025,6.5e5)(2026,5e5)(2026,9.739e3)};

\addplot+[only marks,mark=square*,mark size=2.8pt,color=myGold,mark options={draw=myGold,fill=myGold}]
  coordinates{(2025,3.8581e4)};

\hwach{ibmBlue}{square*}{(2016,5)(2017,17)(2020,65)(2021,127)(2022,433)(2023,1121)}
\hwroad{ibmBlue}{square}{(2023,1121)(2024,1386)(2025,4158)(2033,1e5)}

\hwach{googleRed}{square*}{(2015,9)(2018,72)(2019,53)(2024,105)}
\hwroad{googleRed}{square}{(2024,105)(2030,1e6)}

\hwach{quantGreen}{*}{(2020,6)(2020,10)(2021,12)(2023,32)(2024,56)}
\hwroad{quantGreen}{o}{(2024,56)(2025,100)(2027,1000)(2029,5000)}

\hwach{ionMag}{*}{(2017,5)(2019,11)(2021,23)(2022,32)}
\hwroad{ionMag}{o}{(2022,32)(2025,100)(2027,1e4)(2030,2e6)}

\addplot+[only marks, mark=triangle*, mark size=2.8pt, color=neutralTeal, line width=1pt,
          mark options={draw=neutralTeal,fill=neutralTeal}, forget plot]
  coordinates{(2020,256)(2021,100)(2023,1180)(2025,3000)};
\hwroad{neutralTeal}{triangle}{(2025,3000)(2025,1e4)(2028,5e4)(2030,1e5)}


\hwach{darpaGold}{diamond*}{(2024,16)} 
\hwroad{darpaGold}{diamond}{(2024,16)(2030,2000)} 

\hwach{diraqCyan}{pentagon*}{(2023,2)(2025,10)} 
\hwroad{diraqCyan}{pentagon}{(2025,10)(2029,1e3)(2031,1e6)(2033,1e7)} 


\addlegendimage{only marks, mark=*,mark size=2.8pt, color=algGray, mark options={draw=algGray,fill=algGray}}
\addlegendentry{Alg.\ RSA-2048}
\addlegendimage{only marks, mark=square*,mark size=2.8pt, color=eccBlack, mark options={draw=eccBlack,fill=eccBlack}}
\addlegendentry{Alg.\ ECC-256 (lit.)}
\addlegendimage{only marks, mark=square*,mark size=2.8pt, color=myGold, mark options={draw=myGold,fill=myGold}}
\addlegendentry{Alg.\ ECC-256 (this work)}
\hwlegend{ibmBlue}{square*}
\addlegendentry{IBM}
\hwlegend{googleRed}{square*}
\addlegendentry{Google}
\hwlegend{darpaGold}{diamond*}
\addlegendentry{Alice \& Bob}
\hwlegend{quantGreen}{*}
\addlegendentry{Quantinuum}
\hwlegend{ionMag}{*}
\addlegendentry{IonQ}
\hwlegend{neutralTeal}{triangle*}
\addlegendentry{Neutral-atom}
\hwlegend{diraqCyan}{pentagon*}
\addlegendentry{Diraq}

\end{semilogyaxis}
\end{tikzpicture}
\caption{Algorithmic resource estimates and hardware roadmaps on a common physical-qubit scale. Gray circles: RSA-2048 (algorithms). Black squares: ECC-256 (algorithms). Gold square: ECC-256 (this work). Filled marks indicate achieved results; unfilled marks indicate vendor projections. Algorithmic counts are costed via a 2D surface code with lattice surgery; the cat qubit count from Gouzien \emph{et al.} is shown for comparison \cite{Gouzien2023_ECC}. The 2026 gray circle at $\sim\!10^5$ is the Pinnacle RSA-2048 estimate on a native QLDPC architecture \cite{Webster2026_Pinnacle}; the 2026 black square at $5\times10^5$ is the Babbush \emph{et al.} ECC-256 surface-code estimate \cite{Babbush2026_ECC}. The lowest 2026 gray circle ($\sim\!1.3\times10^4$) and black square ($\sim\!9{,}700$) are the Cain \emph{et al.} balanced RSA-2048 and space-efficient ECC-256 estimates on a reconfigurable neutral-atom architecture with QLDPC codes \cite{Cain2026_Oratomic}. Vendor markers and trajectories draw from public sources: IBM's 2025 roadmap update \cite{IBM_Roadmap2025}; Google's Willow announcements and specification \cite{GoogleRoadmap2023,GoogleBlog2024_Willow,Google2024_Willow105}; Quantinuum H2 updates and roadmap \cite{Quantinuum2024_H2upgrade,QuantinuumRoadmap2024}; IonQ investor roadmap \cite{IonQRoadmap2025}; Alice\&Bob roadmap and funding news \cite{AliceBobRoadmap2024,AliceBobRoadmap2024Long,EETimes2025_AliceBob}; neutral-atom plans (Pasqal, Atom) \cite{NeutralAtomRoadmap2024,Atom2025_3000Q}; and Diraq silicon spin-qubit foundry results and roadmap \cite{DiraqFoundry2025_Nature,DiraqRoadmap2026_GTC}. The shaded band indicates an uncertainty window (2027--2033).}
\label{fig:roadmapsECDLP}
\end{figure}

We compile these developments in Figure~\ref{fig:roadmapsECDLP}. 
To compare results produced years apart, we translate logical resource counts (width, Toffoli/$T$-count, depth) into physical-qubit requirements using a single, widely used costing model:
(i) a 2D surface code with lattice surgery \cite{Litinski2019};
(ii) non-Clifford cost is $T$-limited with dedicated magic-state factories, using Bravyi--Haah distillation through 2023 and magic-state cultivation thereafter \cite{Bravyi2012,Gidney2024_MSCultivation};
(iii) the code distance is chosen from a fixed per-job failure budget; and (iv) we count all tiles, including data, factory, and routing infrastructure. 
One exception in Fig.~\ref{fig:roadmapsECDLP} is the 2023 ECC-256 black square at $126{,}133$, which reports the cat-qubit physical count from Gouzien \emph{et\,al.}~\cite{Gouzien2023_ECC}; we include it as a cross-architecture comparator and do not re-cost it through a 2D surface code. Gidney and Eker{\aa} give the canonical end-to-end example for RSA-2048 under these assumptions \cite{GidneyEkera2019_RSA2048}.
The 2026 Pinnacle RSA-2048 estimate is plotted in the gray series on its native QLDPC architecture \cite{Webster2026_Pinnacle}; the 2026 Babbush \emph{et al.} ECC-256 estimate is plotted as a black square using the common surface-code model \cite{Babbush2026_ECC}.
Cain \emph{et al.} show that combining Gidney's compilation with reconfigurable neutral-atom hardware and QLDPC codes brings RSA-2048 to $\sim\!13{,}000$ atoms and ECC-256 to fewer than $10{,}000$ atoms, albeit at multi-day runtimes on a slow-clock platform \cite{Cain2026_Oratomic}; we plot these in Fig.~\ref{fig:roadmapsECDLP} alongside the other 2026 estimates because architectural breakthroughs that reduce qubit counts on one platform can inform future designs on others.
The same rules are applied when back-casting earlier ECDLP estimates and when placing newer ones.

Three inflection points explain the downward trend in the algorithmic dots.
First, after 2012, practical surface code techniques made logical-to-physical translations possible \cite{fowler2012surface,Fowler2009_SurfaceCodeCosts,jones2012layered,Bravyi2012}.
Second, for ECDLP, concrete logical counts appeared in 2017 and improved in 2020 \cite{Roetteler2017_ECC,Haner2020}. Architectural costing (active volume) and alternative encodings then pushed physical counts lower in 2023 \cite{Litinski2023_ECC,Gouzien2023_ECC}.
Third, in 2024--2026, two distinct threads appeared: direct output compression lowered ECDLP width, while architectural and factory advances drove RSA-2048 from below one million noisy qubits toward the $10^5$ regime \cite{Chevignard2026_ECDLP,Gidney2024_MSCultivation,Gidney2025_RSA2048,Webster2026_Pinnacle}. The former mainly changes width-versus-work tradeoffs, whereas the latter shows how much architecture alone can move the physical count.
Cross-stack studies that sweep hardware parameters provide useful checks \cite{Webber2022,Beverland2022,gheorghiu2019benchmarking}.

On the hardware side, superconducting platforms (IBM, Google) emphasize dense 2D arrays with fast cycles; trapped-ion platforms (Quantinuum, IonQ) trade speed for fidelity and provide native all-to-all connectivity within zones; neutral-atom arrays (Pasqal, QuEra, Atom) scale quickly in qubit count while fault-tolerant demonstrations are still maturing.
A complementary superconducting path is the bosonic \emph{cat-qubit} approach pursued by Alice \& Bob, which leverages strong noise bias to lower physical-to-logical overhead \cite{AliceBobRoadmap2024Long,EETimes2025_AliceBob,AliceBobHelium2023}.
Diraq pursues silicon spin qubits fabricated in a standard CMOS foundry; their 2025 Nature result demonstrates reproducible two-qubit gate fidelities above 99\% on industrially manufactured devices \cite{DiraqFoundry2025_Nature}, and their public roadmap targets $10^7$~physical qubits by 2033 \cite{DiraqRoadmap2026_GTC}.
Public roadmaps show ambitious growth in both qubit count and quality.
These ingredients determine the viable code distance and factory throughput that underlie the dashed lines \cite{IBM2016_5Q,IBM2017_17Q,IBM2020_65Q,IBM2021_Eagle127,IBM2022_Osprey433,IBM2023_Condor1121,IBM_Roadmap2025,Kelly2015_Nature9Q,Google2017_Bristlecone,Arute2019_Sycamore53,GoogleRoadmap2023,Quantinuum2021_H1,Quantinuum2023_H2,Quantinuum2024_H2upgrade,QuantinuumRoadmap2024,Linke2017_PNAS5Q,IonQ2019_11Q,IonQ2021_23Q,IonQ2022_32Q,IonQRoadmap2025,Pasqal2020_256Q,Atom2023_1180Q,Atom2025_3000Q,NeutralAtomRoadmap2024}.

When algorithmic curves and vendor trajectories are overlaid on this common ruler, the earliest intersections appear in the late 2020s; more conservative crossings cluster in the early 2030s.
We therefore indicate a first plausible window for cryptanalytically relevant quantum computers (CRQCs) around 2027--2033.
The endpoints move mainly with three levers: reliable magic-state supply at scale (distillation or cultivation), code distance sufficient for multi-hour jobs, and classical-control latency that keeps pace with fast error-correction cycles.
If any lever stalls, the window shifts to the right; if several improve together, it shifts to the left.

The figure is not a prediction but a calibrated overlay.
A single set of assumptions makes algorithmic progress and hardware progress additive rather than competing narratives, and makes explicit which advances would most accelerate the arrival of CRQCs.

\clearpage
\section{Discussion}
\label{sec:Discussion}

The ECDLP is in BQP, so in principle a sharp classical-to-quantum transition can be expected once sufficiently large and reliable fault-tolerant quantum computers are built.
The challenge suite introduced here converts that abstract statement into an empirically measured phenomenon: a family of $\texttt{secp}k\texttt{k1}$ instances that uses the economically significant Bitcoin curve form while scaling difficulty.
Because the construction is reproducible and the correctness criteria are unambiguous, the same ruler can be used by algorithm designers, hardware teams, and systems engineers without coordination overhead.

The classical record remains consistent with $\Theta(2^{b/2})$ scaling for generic prime-field curves (Section~\ref{sec:ClassicalResources}); constant-factor engineering wins have not changed the asymptotics.
In parallel, logical-to-physical translations still place credible ECC-256 attacks via practical Shor schedules in the mid-$10^{5}$ to low-$10^{6}$ noisy-qubit range under surface-code assumptions, with cat-qubit architectures offering alternative overhead tradeoffs (Section~\ref{sec:QuantumResources}; \cite{Gouzien2023_ECC,Gidney2024_MSCultivation,Gidney2025_RSA2048}) by trading fewer physical qubits for an increased complexity of their architecture.
The newer Chevignard width-optimized ECDLP circuit and the Pinnacle RSA architecture sharpen opposite sides of the design space: the former cuts width but raises total non-Clifford work by orders of magnitude, while the latter shows that a favorable QLDPC architecture can slash factoring qubit counts without yet yielding a directly comparable ECC-256 estimate \cite{Chevignard2026_ECDLP,Webster2026_Pinnacle}.
Overlaying algorithmic cost with public roadmaps yields a first plausible window for cryptanalytically relevant quantum computers in roughly 2027--2033, albeit with wide error bars.

Our resource counts inherit all the usual caveats: they depend on cycle times, physical error rates, code distances, factory throughput and layout, decoding latency, and how aggressively one amortizes failure budgets.
They also assume particular algorithmic decompositions. For example, better phase-gradient synthesis and cultivation can shrink the non-Clifford footprint; compression-robust period finding and Legendre-symbol output compression can shift width-versus-work tradeoffs; and active-volume tricks such as batch inversion or state reuse can reduce depth in specific layouts \cite{Haner2020,Litinski2023_ECC,MaySchlieper2022,Gidney2024_MSCultivation,Chevignard2026_ECDLP}.
On the hardware side, noise bias and bosonic encodings change the physical-to-logical exchange rate in ways not captured by a vanilla 2D surface code \cite{Gouzien2023_ECC,Ruiz2025_UnfoldedDistillation}.
Consequently, the dots in Fig.~\ref{fig:roadmapsECDLP} should be read as order-of-magnitude waypoints, not promises.

For Bitcoin, the operational risk centers on addresses whose public keys have been (or will be) revealed on chain: once a public key is visible, an adversary with a CRQC can, in principle, derive the corresponding private key fast enough to race in the mempool or drain exposed UTXOs \cite{Aggarwal2018}.
Macro-level analyses suggest that even the credible prospect of such capability has market-design consequences \cite{Rohde2021_QCryptoEcon}.
Any responsible migration therefore needs (i) a plan to move value off legacy outputs before the risk window opens, and (ii) script-level mechanisms that let participants adopt quantum-safe verification paths without hard forks or excessive coordination.
Concretely, BIP~360 proposes introducing pay-to-quantum-resistant-hash (\texttt{P2QRH}) output types via a soft fork, providing PQC-ready address semantics that migration plans such as Post-Quantum Migration can leverage \cite{BIP360,Lopp2025_BIP_PQMig}.

Two broad approaches are complementary rather than exclusive.
First, \emph{slow-defence} and \emph{commit-then-upgrade} techniques create opt-in timelines that give the network room to react if the risk profile changes suddenly \cite{Stewart2018SlowDefenceBitcoin,Ilie2020CommittingBetter}.
Second, \emph{hybrid signatures} combine classical and post-quantum verification, degrading gracefully if one primitive is later weakened.
For Bitcoin specifically, careful use of Taproot's policy flexibility can embed migration levers without breaking existing spend conditions.
Operationally, recent work argues that some amount of coordinated downtime (or at least staged quiet periods) may be unavoidable to sweep long-tail UTXOs \cite{pont2024downtime}.
These ideas are not mutually exclusive and can be staged: commit today, migrate when safe and convenient, and retain hybrid fallbacks while PQC matures.
A concrete migration path for Bitcoin follows a commit-then-upgrade pattern.
Coins first move to a script that commits, via a hash, to a post-quantum (PQ) public key, or to a Merkle root of PQ keys, while preserving a standard ECDSA/Taproot spend branch; no on-chain PQ verification is required at this stage.
During the transition, users spend normally through the classical branch while public keys remain unrevealed on chain; because Shor's algorithm targets revealed public keys, this keeps dormant UTXOs protected.
Once a soft fork introduces PQ verification (for example, a \texttt{PQSIGVERIFY}-style opcode for a NIST-selected signature), the committed PQ branch becomes enforceable and funds can be swept under PQ-only control; timelocks can be used to desynchronize the reveal and the spend.
This approach fits the slow-defence literature and aligns with recent migration proposals that pair PQ activation with a phased sunset of legacy signatures \cite{Stewart2018SlowDefenceBitcoin,Ilie2020CommittingBetter,Lopp2025_BIP_PQMig}.

A credible demonstration of a nontrivial rung on the ladder coupled with transparent reporting of logical- and physical-level metrics would constitute a de-risking event for migration planning: the shift from dozens to thousands of logical qubits within a short horizon is the relevant threshold for ECC. 
Because the mapping from hardware roadmaps to ECDLP runtime is now explicit, the economically prudent course is to complete upgrades to post-quantum signatures before early fault-tolerant devices reach the 160-256-bit rungs.

Taken together, the challenge ladder, classical calibration, and quantum costing give a shared language for a community that rarely speaks together. Our view is deliberately conservative: it is easier to be surprised on the upside (better algorithms, more qubits, or both) than to compress the exponential wall of Fig.~\ref{fig:classicalECDLPruntime}.
Either way, preparing migration levers now is cheap insurance.

\clearpage
\bibliographystyle{quantum}
\bibliography{references}
\clearpage


\appendix


\section{The list of ECDLP challenges \label{sec:ECDLPchallenges}}

Each card lists the field prime $p$, the prime group order $n$, and the compressed SEC1 NUMS points
$P$ and $Q$ derived from the fixed labels \texttt{"Quantum"} and \texttt{"Challenge"}.
The task in every case is to recover $\kappa$ such that $P=[\kappa]Q$.

\ECCDLcard{6}{43}{0307}{0320}{31}
\ECCDLcard{8}{163}{0205}{0261}{139}
\ECCDLcard{12}{2089}{0207C0}{03053E}{2143}
\ECCDLcard{16}{32803}{0260A3}{036D88}{32497}
\ECCDLcard{24}{16777213}{0276BA79}{0385303B}{16770451}
\ECCDLcard{32}{4294966177}{021E0EF3EB}{02EF01BFF4}{4294835173}
\ECCDLcard{48}{281474976707983}{03769EC7F35857}{02210644F01A41}{281474961723409}
\ECCDLcard{64}{18446744073709546873}{037F3EBCB70E7FF266}{03989C2061EF0E994E}{18446744079408176167}
\ECCDLcard{80}{1208925819614629174697917}{030D08430A6C00FAE05651}{02CF45D656507F3699875E}{1208925819612459780337609}
\ECCDLcard{96}{79228162514264337593543943091}{029E1F276C17BCE3D8DF9D22BB}{03CCF37E1E144A24433C0582A6}{79228162514264890447722092359}
\ECCDLcard{112}{5192296858534827628530496329219907}{02949E2914E4863781500B38122BFD}{03EAE0A2B0C8B159B55FA7C5BF18D0}{5192296858534827704501506384124197}
\ECCDLcard{128}{340282366920938463463374607431768193873}{02429EC9073D11651ACFA5293FC7641CE5}{03871B2CDD66580C22D129B3DBA13AADE4}{340282366920938463493968231006123201901}
\ECCDLcard{144}{22300745198530623141535718272648361505978547}{03635AC435AB7D9405A142D55723D4588B2392}{03B33F4F69A88768A79D3F029C3928D61FF18D}{22300745198530623141528449602726112266830289}
\ECCDLcard{160}{1461501637330902918203684832716283019655932523797}{022E9A635AC20A338BCF78452EE06D87F74BAB9888}{02A15CB33F4E47113EB4981A4E92BD658CD5307B1D}{1461501637330902918203683111413834986743050038629}
\ECCDLcard{176}{95780971304118053647396689196894323976171195136464057}{0370A72E9A635AC20A1D46F3921D13BB3673C1B54BA654}{037BB4A15CB33F4E4705985367CE2E3EB95382ABA25799}{95780971304118053647396688993621402904752667866498983}
\ECCDLcard{192}{6277101735386680763835789423207666416102355444464034510271}{02362870A72E9A635AC20A1D46E6B4EBB7526089BC79F0B9D2}{0355117BB4A15CB33F4E4705984CACFC6137DAE282E7F936CA}{6277101735386680763835789423049239630206963053826514709771}
\ECCDLcard{208}{411376139330301510538742295639337626245683966408394965837126831}{023A98362870A72E9A635AC20A1D46E6B4E8AB299942EB67C20297}{024DA955117BB4A15CB33F4E4705984CACFAC914CBD728FBAF980D}{411376139330301510538742295639309734542636196754601463706467431}
\ECCDLcard{224}{26959946667150639794667015087019630673637144422540572481103610247941}{02549C3A98362870A72E9A635AC20A1D46E6B4E8AB0C13C010E18EEB2D}{03F7BB4DA955117BB4A15CB33F4E4705984CACFAC9055A294AF4CC490A}{26959946667150639794667015087019637901048645921160586259561071496361}
\ECCDLcard{240}{1766847064778384329583297500742918515827483896875618958121606201292517917}{036903549C3A98362870A72E9A635AC20A1D46E6B4E8AB0C13BE965A409A82}{021792F7BB4DA955117BB4A15CB33F4E4705984CACFAC9055A2884EE41346F}{1766847064778384329583297500742918517618513573386834281239873043187462749}
\ECCDLcard{256}{115792089237316195423570985008687907853269984665640564039457584007908834671663}{034C186903549C3A98362870A72E9A635AC20A1D46E6B4E8AB0C13BE95E3FBE93A}{0327CF1792F7BB4DA955117BB4A15CB33F4E4705984CACFAC9055A2884B061E4E3}{115792089237316195423570985008687907852837564279074904382605163141518161494337}

\clearpage


\begin{landscape}
\section{Full resource estimation data for Sec.\ref{sec:QuantumResources} \label{sec:FullQuantumResources}}

\begin{table}[ht]
  \centering
  \begin{tabular}{S[table-figures-integer=3]S[table-figures-integer=3]|S[table-figures-integer=2]S[table-figures-integer=1]S[table-figures-integer=2]S[table-figures-integer=2]S[table-figures-integer=2]|S[table-figures-integer=2]S[table-figures-integer=5]S[table-figures-integer=6]ccS[table-figures-integer=4]}
  {$k$}                         &{$w_{\mathrm{exp}}$}          &{$w_e$}                       &{$w_m$}                       &{$\Lambda$}                   &{$D$}                         &{$i$}                         &{$\#\text{factories}$}        &{$N_{\mathrm{factory}}$}      &{$N_{\mathrm{phys}}$}         &{$t$}                         &{$t_{\mathrm{exp}}$}          &{$N_{\mathrm{log}}$} \\
        6                             &12                            &9                             &2                             &12                            &7                             &4                             &5                             &388                           &2311                          &\SI{513}{\milli\second}       &\SI{545}{\milli\second}       &67                            \\
        8                             &16                            &9                             &2                             &12                            &7                             &4                             &5                             &388                           &2824                          &\SI{1}{\second}               &\SI{1}{\second}               &85                            \\
        12                            &24                            &10                            &3                             &13                            &7                             &4                             &5                             &388                           &3869                          &\SI{3}{\second}               &\SI{4}{\second}                &122                           \\
        16                            &32                            &11                            &4                             &14                            &9                             &5                             &6                             &463                           &5968                          &\SI{9}{\second}               &\SI{10}{\second}              &159                           \\
        24                            &48                            &12                            &4                             &15                            &9                             &6                             &12                            &1157                          &9165                          &\SI{25}{\second}              &\SI{28}{\second}               &232                           \\
        32                            &64                            &13                            &4                             &15                            &9                             &7                             &16                            &1537                          &12075                         &\SI{55}{\second}              &\SI{1}{\minute}               &305                           \\
        48                            &96                            &14                            &4                             &16                            &11                            &7                             &13                            &1252                          &19485                         &\SI{3}{\minute}               &\SI{4}{\minute}               &450                           \\
        64                            &128                           &15                            &4                             &17                            &11                            &7                             &13                            &1252                          &25371                         &\SI{8}{\minute}               &\SI{9}{\minute}               &595                           \\
        80                            &160                           &15                            &5                             &17                            &11                            &8                             &20                            &1917                          &31868                         &\SI{14}{\minute}              &\SI{17}{\minute}              &739                           \\
        96                            &192                           &16                            &6                             &17                            &11                            &8                             &20                            &1917                          &37727                         &\SI{24}{\minute}              &\SI{34}{\minute}              &884                           \\
        112                           &224                           &16                            &6                             &18                            &13                            &9                             &52                            &6001                          &53811                         &\SI{43}{\minute}              &\SI{48}{\minute}              &1028                          \\
        128                           &256                           &17                            &5                             &18                            &13                            &10                            &87                            &10026                         &64594                         &\SI{1}{hour}                  &\SI{1}{hour}                  &1173                          \\
        144                           &288                           &17                            &6                             &19                            &13                            &10                            &87                            &10026                         &71290                         &\SI{1}{hour}                  &\SI{2}{hours}                 &1317                          \\
        160                           &320                           &17                            &6                             &19                            &13                            &10                            &87                            &10026                         &77986                         &\SI{2}{hours}                 &\SI{2}{hours}                 &1461                          \\
        176                           &352                           &17                            &6                             &19                            &13                            &10                            &87                            &10026                         &84682                         &\SI{3}{hours}                 &\SI{3}{hours}                 &1605                          \\
        192                           &384                           &18                            &6                             &19                            &13                            &10                            &87                            &10026                         &91409                         &\SI{3}{hours}                 &\SI{4}{hours}                 &1750                          \\
        208                           &416                           &18                            &6                             &19                            &13                            &10                            &87                            &10026                         &98105                         &\SI{4}{hours}                 &\SI{5}{hours}                 &1894                          \\
        224                           &448                           &18                            &6                             &19                            &13                            &10                            &87                            &10026                         &104801                        &\SI{5}{hours}                 &\SI{6}{hours}                 &2038                          \\
        240                           &480                           &18                            &6                             &19                            &13                            &10                            &87                            &10026                         &111497                        &\SI{6}{hours}                 &\SI{8}{hours}                 &2182                          \\
        256                           &512                           &18                            &6                             &19                            &13                            &12                            &84                            &18101                         &126260                        &\SI{7}{hours}                 &\SI{9}{hours}                 &2326                          \\
\end{tabular}
  \caption{Nominal size-based ECDLP resource estimates using the repetition cat code architecture Alice \& Bob \cite{Gouzien2023_ECC}.
  The columns are, in order, the nominal field bit length $k$; the total exponent bit count $w_{\mathrm{exp}}=2k$ used by this model (the approximation $b\approx k$, not an exact per-card control width); the exponent window size $w_e$ in bits; the multiplier or Montgomery window size $w_m$ in bits used in point-arithmetic lookups; the mean photon number per cat $\Lambda \equiv \alpha^2$, which sets the noise bias; the repetition cat code distance $D$; the number of QEC correction cycles per logical time step $i$; the number of concurrent magic-state factories $\#\text{factories}$; the total physical qubits allocated to all factories $N_{\mathrm{factory}}$; the total physical qubits $N_{\mathrm{phys}}$ (data plus factories); the critical-path runtime $t$; the expected runtime including repetitions and overheads $t_{\mathrm{exp}}$; and the total logical qubits $N_{\mathrm{log}}$.
  The data was obtained using the code in \cite{Gouzien2024_elliptic_log_cat}.}
\end{table}
\clearpage

\begin{table}[ht]
  \centering
  \begin{tabular}{S[table-figures-integer=3]S[table-figures-integer=3]|S[table-figures-integer=2]S[table-figures-integer=1]S[table-figures-integer=2]S[table-figures-integer=2]S[table-figures-integer=2]|S[table-figures-integer=2]S[table-figures-integer=4]S[table-figures-integer=7]ccS[table-figures-integer=4]}
  {$k$}                         &{$w_{\mathrm{exp}}$}          &{$w_e$}                       &{$w_m$}                       &{$\Lambda$}                   &{$D$}                         &{$i$}                         &{$\#\text{factories}$}        &{$N_{\mathrm{factory}}$}      &{$N_{\mathrm{phys}}$}         &{$t$}                         &{$t_{\mathrm{exp}}$}          &{$N_{\mathrm{log}}$}  \\
        6                             &12                            &9                             &2                             &21                            &22                            &5                             &2                             &94                            &1194                        &\SI{2}{\second}               &\SI{2}{\second}               &68                            \\
        8                             &16                            &10                            &2                             &21                            &22                            &5                             &2                             &94                            &1455                        &\SI{5}{\second}               &\SI{5}{\second}               &86                            \\
        12                            &24                            &10                            &3                             &21                            &22                            &5                             &2                             &94                            &1977                        &\SI{14}{\second}     &\SI{14}{\second}              &122                           \\
        16                            &32                            &11                            &4                             &21                            &22                            &6                             &4                             &252                           &2686                        &\SI{29}{\second}              &\SI{29}{\second}              &160                           \\
        24                            &48                            &12                            &4                             &21                            &22                            &7                             &5                             &315                           &3793                        &\SI{1}{\minute}             &\SI{1}{\minute}                &232                           \\
        32                            &64                            &13                            &4                             &21                            &22                            &7                             &5                             &315                           &4866                        &\SI{3}{\minute}               &\SI{3}{\minute}               &306                           \\
        48                            &96                            &14                            &4                             &21                            &22                            &7                             &5                             &315                           &6954                        &\SI{10}{\minute}              &\SI{10}{\minute}              &450                           \\
        64                            &128                           &15                            &4                             &21                            &22                            &8                             &8                             &504                           &9260                        &\SI{21}{\minute}              &\SI{22}{\minute}              &596                           \\
        80                            &160                           &15                            &5                             &21                            &22                            &8                             &8                             &504                           &11348                       &\SI{40}{\minute}              &\SI{42}{\minute}              &740                           \\
        96                            &192                           &16                            &6                             &21                            &22                            &8                             &8                             &504                           &13436                       &\SI{1}{hours}                 &\SI{1}{hours}                 &884                           \\
        112                           &224                           &16                            &6                             &21                            &22                            &10                            &37                            &2923                          &17943                       &\SI{2}{hours}                 &\SI{2}{hours}                 &1028                          \\
        128                           &256                           &17                            &5                             &21                            &22                            &10                            &37                            &2923                          &20060                       &\SI{2}{hours}                 &\SI{2}{hours}                 &1174                          \\
        144                           &288                           &17                            &6                             &21                            &22                            &10                            &37                            &2923                          &22148                       &\SI{3}{hours}                 &\SI{3}{hours}                 &1318                          \\
        160                           &320                           &17                            &6                             &21                            &22                            &10                            &37                            &2923                          &24236                       &\SI{5}{hours}                 &\SI{5}{hours}                 &1462                          \\
        176                           &352                           &17                            &6                             &21                            &22                            &10                            &37                            &2923                          &26324                       &\SI{6}{hours}                 &\SI{6}{hours}                 &1606                          \\
        192                           &384                           &18                            &6                             &21                            &22                            &10                            &37                            &2923                          &28412                       &\SI{8}{hours}                 &\SI{8}{hours}                 &1750                          \\
        208                           &416                           &18                            &6                             &21                            &22                            &10                            &37                            &2923                          &30500                       &\SI{10}{hours}                &\SI{10}{hours}                &1894                          \\
        224                           &448                           &18                            &6                             &21                            &22                            &10                            &37                            &2923                          &32588                       &\SI{12}{hours}                &\SI{13}{hours}                &2038                          \\
        240                           &480                           &18                            &6                             &21                            &22                            &10                            &37                            &2923                          &34676                       &\SI{14}{hours}                &\SI{16}{hours}                &2182                          \\
        256                           &512                           &18                            &6                             &21                            &22                            &11                            &60                            &4740                          &38581                       &\SI{17}{hours}                &\SI{18}{hours}                &2326                          \\
\end{tabular}
  \caption{Nominal size-based ECDLP resource estimates using the LDPC cat code architecture Alice \& Bob \cite{ruiz2025ldpc}.
  The columns are, in order, the nominal field bit length $k$; the total exponent bit count $w_{\mathrm{exp}}=2k$ used by this model (the approximation $b\approx k$, not an exact per-card control width); the exponent window size $w_e$ in bits; the multiplier or Montgomery window size $w_m$ in bits used in point-arithmetic lookups; the mean photon number per cat $\Lambda \equiv \alpha^2$, which sets the noise bias; the LDPC cat code distance $D$; the number of QEC correction cycles per logical time step $i$; the number of concurrent magic-state factories $\#\text{factories}$; the total physical qubits allocated to all factories $N_{\mathrm{factory}}$; the total physical qubits $N_{\mathrm{phys}}$ (data plus factories); the critical-path runtime $t$; the expected runtime including repetitions and overheads $t_{\mathrm{exp}}$; and the total logical qubits $N_{\mathrm{log}}$.
  The data was obtained using the code in \cite{Gouzien2024_elliptic_log_cat}.}
\end{table}
\clearpage
\end{landscape}

\clearpage
\begin{landscape}
\input{resource_estimator/table_logical_lowD_fixed.tex}
\end{landscape}

\clearpage
\begin{landscape}
\input{resource_estimator/table_logical_lowT_fixed.tex}
\end{landscape}

\clearpage
\begin{landscape}
\input{resource_estimator/table_logical_lowW_fixed.tex}
\end{landscape}

\clearpage
\begin{landscape}
\input{resource_estimator/table_phys_lowD_aggressive.tex}
\end{landscape}

\clearpage
\begin{landscape}
\input{resource_estimator/table_phys_lowD_conservative.tex}
\end{landscape}

\clearpage
\begin{landscape}
\input{resource_estimator/table_phys_lowT_aggressive.tex}
\end{landscape}

\clearpage
\begin{landscape}
\input{resource_estimator/table_phys_lowT_conservative.tex}
\end{landscape}

\clearpage
\begin{landscape}
\input{resource_estimator/table_phys_lowW_aggressive.tex}
\end{landscape}

\clearpage
\begin{landscape}
\input{resource_estimator/table_phys_lowW_conservative.tex}
\end{landscape}

\end{document}

%% file: bitcoin-logo.pdf_tex
\begingroup%
  \makeatletter%
  \providecommand\color[2][]{%
    \errmessage{(Inkscape) Color is used for the text in Inkscape, but the package 'color.sty' is not loaded}%
    \renewcommand\color[2][]{}%
  }%
  \providecommand\transparent[1]{%
    \errmessage{(Inkscape) Transparency is used (non-zero) for the text in Inkscape, but the package 'transparent.sty' is not loaded}%
    \renewcommand\transparent[1]{}%
  }%
  \providecommand\rotatebox[2]{#2}%
  \newcommand*\fsize{\dimexpr\f@size pt\relax}%
  \newcommand*\lineheight[1]{\fontsize{\fsize}{#1\fsize}\selectfont}%
  \ifx\svgwidth\undefined%
    \setlength{\unitlength}{3068.4525bp}%
    \ifx\svgscale\undefined%
      \relax%
    \else%
      \setlength{\unitlength}{\unitlength * \real{\svgscale}}%
    \fi%
  \else%
    \setlength{\unitlength}{\svgwidth}%
  \fi%
  \global\let\svgwidth\undefined%
  \global\let\svgscale\undefined%
  \makeatother%
  \begin{picture}(1,1.00011243)%
    \lineheight{1}%
    \setlength\tabcolsep{0pt}%
    \put(0,0){\includegraphics[width=\unitlength,page=1]{bitcoin-logo.pdf}}%
  \end{picture}%
\endgroup%

%% file: resource_estimator/table_logical_lowD_fixed.tex
\begin{center}
\small
\setlength{\tabcolsep}{5pt}
\renewcommand{\arraystretch}{1.05}
\captionof{table}{Microsoft Resource Estimator logical totals for Shor ECDLP (low-depth-optimal schedule).}
\label{tab:ms-logical-low-depth}
\resizebox{\linewidth}{!}{%
\begin{tabular}{@{}r S[table-format=2.0] S[table-format=11.0] S[table-format=10.0] S[table-format=9.0] S[table-format=10.0] S[table-format=9.0] S[table-format=10.0] S[table-format=4.0]@{}}
\toprule
\multicolumn{1}{c}{Bits} & \multicolumn{1}{c}{w} & \multicolumn{1}{c}{CNOT} & \multicolumn{1}{c}{1q} & \multicolumn{1}{c}{Meas.} & \multicolumn{1}{c}{T} & \multicolumn{1}{c}{T-depth} & \multicolumn{1}{c}{Full depth} & \multicolumn{1}{c}{Width} \\
\midrule
6 & 3 & 259120 & 164232 & 8657 & 359272 & 66804 & 344390 & 83 \\
8 & 4 & 1171760 & 368384 & 28644 & 699632 & 120598 & 775034 & 102 \\
12 & 6 & 3317433 & 806138 & 68610 & 1533528 & 247654 & 1623404 & 152 \\
16 & 8 & 6822500 & 1495036 & 135404 & 2803768 & 437760 & 2929320 & 202 \\
24 & 12 & 25514568 & 4037727 & 438475 & 6608608 & 996672 & 7831138 & 300 \\
32 & 16 & 201518153 & 18154085 & 2917380 & 17507104 & 2485460 & 37667343 & 399 \\
48 & 16 & 353566947 & 38132947 & 5377919 & 46937844 & 6762858 & 76342091 & 583 \\
64 & 16 & 567904716 & 71635026 & 9096095 & 102028400 & 14796496 & 139390846 & 760 \\
80 & 16 & 863456806 & 122686752 & 14434669 & 190387160 & 27672540 & 233945868 & 935 \\
96 & 16 & 1261741064 & 196089999 & 21866243 & 321275280 & 46755480 & 368491714 & 1111 \\
112 & 16 & 1782336329 & 295799423 & 31723130 & 501995340 & 73071460 & 549945995 & 1286 \\
128 & 16 & 2452784669 & 428790788 & 44756718 & 745509824 & 108618448 & 791826221 & 1464 \\
144 & 18 & 5051129831 & 676745746 & 82751236 & 1012119456 & 146034429 & 1293177486 & 1645 \\
160 & 18 & 6048400100 & 865164420 & 100827920 & 1361856276 & 196748605 & 1636092986 & 1822 \\
176 & 18 & 7415080217 & 1113338450 & 125881679 & 1805086480 & 261055620 & 2088627522 & 1997 \\
192 & 18 & 9064818690 & 1414700418 & 156300100 & 2343601220 & 339279894 & 2637488728 & 2174 \\
208 & 19 & 13958247102 & 1885854071 & 227479292 & 2865540864 & 412543714 & 3584950952 & 2351 \\
224 & 19 & 16087646111 & 2276801054 & 266648374 & 3569725632 & 514684140 & 4296269609 & 2527 \\
240 & 20 & 26624934518 & 3165821773 & 413692601 & 4379276688 & 626175051 & 6132009701 & 2705 \\
256 & 20 & 28959031250 & 3644646292 & 459043525 & 5276241436 & 756915745 & 6992537566 & 2884 \\
\bottomrule
\end{tabular}%
}
\end{center}

%% file: resource_estimator/table_logical_lowT_fixed.tex
\begin{center}
\small
\setlength{\tabcolsep}{5pt}
\renewcommand{\arraystretch}{1.05}
\captionof{table}{Microsoft Resource Estimator logical totals for Shor ECDLP (low-T-optimal schedule).}
\label{tab:ms-logical-low-t}
\resizebox{\linewidth}{!}{%
\begin{tabular}{@{}r S[table-format=2.0] S[table-format=11.0] S[table-format=10.0] S[table-format=9.0] S[table-format=10.0] S[table-format=9.0] S[table-format=10.0] S[table-format=4.0]@{}}
\toprule
\multicolumn{1}{c}{Bits} & \multicolumn{1}{c}{w} & \multicolumn{1}{c}{CNOT} & \multicolumn{1}{c}{1q} & \multicolumn{1}{c}{Meas.} & \multicolumn{1}{c}{T} & \multicolumn{1}{c}{T-depth} & \multicolumn{1}{c}{Full depth} & \multicolumn{1}{c}{Width} \\
\midrule
6 & 3 & 24931 & 101540 & 6442 & 195488 & 87444 & 302398 & 80 \\
8 & 4 & 215611 & 227590 & 20123 & 366936 & 153478 & 639634 & 93 \\
12 & 6 & 1211505 & 472746 & 44398 & 781128 & 312314 & 1310188 & 138 \\
16 & 8 & 2897896 & 833852 & 82044 & 1370872 & 533232 & 2265124 & 184 \\
24 & 12 & 16987605 & 2427030 & 302032 & 3317080 & 1182736 & 6255214 & 275 \\
32 & 11 & 22715283 & 4819410 & 493375 & 7940148 & 2985143 & 12880141 & 353 \\
48 & 12 & 69790930 & 14077653 & 1447553 & 23410288 & 8707239 & 37577613 & 516 \\
64 & 13 & 155651941 & 30877872 & 3173623 & 51639052 & 19118454 & 82123936 & 679 \\
80 & 16 & 615998934 & 72831862 & 9770940 & 93142380 & 31237410 & 182158912 & 847 \\
96 & 16 & 838484861 & 109355738 & 13704936 & 152101800 & 52615020 & 278225126 & 1007 \\
112 & 16 & 1102817260 & 157454811 & 18677541 & 232843492 & 82155766 & 405419577 & 1167 \\
128 & 16 & 1420901238 & 219477827 & 24901281 & 339371488 & 121402416 & 570251987 & 1328 \\
144 & 16 & 1807086247 & 296972320 & 32506021 & 474731964 & 171411804 & 777796744 & 1488 \\
160 & 16 & 2264480848 & 392016924 & 41684361 & 642795480 & 233648140 & 1032088344 & 1648 \\
176 & 16 & 2803006611 & 506706512 & 52637069 & 847289652 & 309493536 & 1339149882 & 1808 \\
192 & 16 & 3432147754 & 642524285 & 65472433 & 1091572176 & 400240728 & 1704197069 & 1968 \\
208 & 18 & 6730490468 & 935081864 & 112052863 & 1373989200 & 481019036 & 2397057497 & 2134 \\
224 & 18 & 7797424079 & 1129112405 & 131980974 & 1701999116 & 600748568 & 2909262635 & 2294 \\
240 & 18 & 8967846469 & 1349475174 & 154296017 & 2079599368 & 739011267 & 3493715381 & 2454 \\
256 & 19 & 14708029780 & 1809016615 & 228485292 & 2507025032 & 858685991 & 4579925011 & 2617 \\
\bottomrule
\end{tabular}%
}
\end{center}

%% file: resource_estimator/table_logical_lowW_fixed.tex
\begin{center}
\small
\setlength{\tabcolsep}{5pt}
\renewcommand{\arraystretch}{1.05}
\captionof{table}{Microsoft Resource Estimator logical totals for Shor ECDLP (low-width-optimal schedule).}
\label{tab:ms-logical-low-width}
\resizebox{\linewidth}{!}{%
\begin{tabular}{@{}r S[table-format=1.0] S[table-format=12.0] S[table-format=11.0] S[table-format=8.0] S[table-format=12.0] S[table-format=11.0] S[table-format=12.0] S[table-format=4.0]@{}}
\toprule
\multicolumn{1}{c}{Bits} & \multicolumn{1}{c}{w} & \multicolumn{1}{c}{CNOT} & \multicolumn{1}{c}{1q} & \multicolumn{1}{c}{Meas.} & \multicolumn{1}{c}{T} & \multicolumn{1}{c}{T-depth} & \multicolumn{1}{c}{Full depth} & \multicolumn{1}{c}{Width} \\
\midrule
6 & 1 & 25808 & 368886 & 1247 & 1219248 & 681502 & 1985088 & 73 \\
8 & 1 & 2074567 & 1051079 & 2047 & 2923648 & 1549213 & 4928927 & 81 \\
12 & 1 & 12813981 & 3727811 & 4222 & 10576320 & 5714012 & 18396215 & 104 \\
16 & 1 & 32755307 & 8597519 & 11956 & 24381504 & 12862075 & 41735519 & 140 \\
24 & 1 & 135747085 & 33036310 & 45582 & 91887936 & 49043944 & 162011278 & 204 \\
32 & 1 & 321370892 & 75455454 & 187239 & 213387776 & 112164661 & 370730941 & 272 \\
48 & 1 & 1201971365 & 278762925 & 464411 & 782709504 & 417236048 & 1386923804 & 400 \\
64 & 1 & 2838639860 & 652926908 & 874191 & 1847084544 & 974006378 & 3245111419 & 529 \\
80 & 1 & 5766317242 & 1328830475 & 2486659 & 3730328320 & 1951019013 & 6589083354 & 660 \\
96 & 1 & 10618079542 & 2419619226 & 3696694 & 6792749568 & 3618861471 & 12202606456 & 788 \\
112 & 1 & 16798262986 & 3837625225 & 4938218 & 10804354624 & 5705909402 & 19274921335 & 916 \\
128 & 1 & 25174658388 & 5718332792 & 6542750 & 16169045504 & 8533547732 & 28814259958 & 1045 \\
144 & 1 & 36202653269 & 8295203367 & 8429649 & 23423467968 & 12284917007 & 41544572820 & 1173 \\
160 & 1 & 50283795853 & 11530473750 & 10259717 & 32606983040 & 17102545545 & 57765267187 & 1301 \\
176 & 1 & 69161078939 & 15848184869 & 12521913 & 44707681920 & 23622083172 & 79757333906 & 1429 \\
192 & 1 & 92597043361 & 21164220852 & 15085677 & 59453297664 & 31693335486 & 107113902705 & 1557 \\
208 & 1 & 119775348413 & 26939326115 & 30500128 & 75674664832 & 40084968281 & 138130068111 & 1688 \\
224 & 1 & 149965334160 & 33647194322 & 35978708 & 94707940096 & 50064207795 & 172786673006 & 1816 \\
240 & 1 & 184316860474 & 41336535585 & 41002376 & 116556830400 & 61530475118 & 212240142701 & 1944 \\
256 & 1 & 224673412698 & 50291020016 & 46455611 & 141721438208 & 74902098856 & 258487044076 & 2073 \\
\bottomrule
\end{tabular}%
}
\end{center}

%% file: resource_estimator/table_phys_lowD_aggressive.tex
\begin{center}
\small
\setlength{\tabcolsep}{5pt}
\renewcommand{\arraystretch}{1.05}
\captionof{table}{Fault-tolerant resources for Shor ECDLP (low-depth-optimal), surface-code aggressive scenario.}
\label{tab:phys-low-depth-aggressive}
\resizebox{\linewidth}{!}{%
\begin{tabular}{@{}r S[table-format=4.0] S[table-format=2.0] S[table-format=1.0] S[table-format=10.0] S[table-format=9.0] S[table-format=10.0] S[table-format=6.0] S[table-format=5.2] S[table-format=2.2]@{}}
\toprule
\multicolumn{1}{c}{Bits} & \multicolumn{1}{c}{LogicalQ} & \multicolumn{1}{c}{d} & \multicolumn{1}{c}{Factories} & \multicolumn{1}{c}{T} & \multicolumn{1}{c}{T-depth} & \multicolumn{1}{c}{Full depth} & \multicolumn{1}{c}{PhysicalQ} & \multicolumn{1}{c}{Time [s]} & \multicolumn{1}{c}{Time [h]} \\
\midrule
6 & 83 & 9 & 6 & 359272 & 66804 & 344390 & 16439 & 2.73 & 0.00 \\
8 & 102 & 9 & 6 & 699632 & 120598 & 775034 & 18978 & 6.14 & 0.00 \\
12 & 152 & 9 & 6 & 1533528 & 247654 & 1623404 & 25661 & 12.86 & 0.00 \\
16 & 202 & 9 & 6 & 2803768 & 437760 & 2929320 & 32343 & 23.20 & 0.01 \\
24 & 300 & 11 & 5 & 6608608 & 996672 & 7831138 & 66550 & 75.81 & 0.02 \\
32 & 399 & 11 & 3 & 17507104 & 2485460 & 37667343 & 83653 & 364.62 & 0.10 \\
48 & 583 & 11 & 4 & 46937844 & 6762858 & 76342091 & 121720 & 738.99 & 0.21 \\
64 & 760 & 11 & 5 & 102028400 & 14796496 & 139390846 & 158389 & 1349.30 & 0.37 \\
80 & 935 & 11 & 5 & 190387160 & 27672540 & 233945868 & 193328 & 2264.60 & 0.63 \\
96 & 1111 & 13 & 5 & 321275280 & 46755480 & 368491714 & 319097 & 4215.55 & 1.17 \\
112 & 1286 & 13 & 6 & 501995340 & 73071460 & 549945995 & 369755 & 6291.38 & 1.75 \\
128 & 1464 & 13 & 6 & 745509824 & 108618448 & 791826221 & 419390 & 9058.49 & 2.52 \\
144 & 1645 & 13 & 5 & 1012119456 & 146034429 & 1293177486 & 468003 & 14793.95 & 4.11 \\
160 & 1822 & 13 & 5 & 1361856276 & 196748605 & 1636092986 & 517360 & 18716.90 & 5.20 \\
176 & 1997 & 13 & 5 & 1805086480 & 261055620 & 2088627522 & 566158 & 23893.90 & 6.64 \\
192 & 2174 & 13 & 6 & 2343601220 & 339279894 & 2637488728 & 617374 & 30172.87 & 8.38 \\
208 & 2351 & 13 & 5 & 2865540864 & 412543714 & 3584950952 & 664871 & 41011.84 & 11.39 \\
224 & 2527 & 13 & 5 & 3569725632 & 514684140 & 4296269609 & 713949 & 49149.32 & 13.65 \\
240 & 2705 & 13 & 5 & 4379276688 & 626175051 & 6132009701 & 763584 & 70150.19 & 19.49 \\
256 & 2884 & 13 & 5 & 5276241436 & 756915745 & 6992537566 & 813498 & 79994.63 & 22.22 \\
\bottomrule
\end{tabular}%
}
\end{center}

%% file: resource_estimator/table_phys_lowD_conservative.tex
\begin{center}
\small
\setlength{\tabcolsep}{5pt}
\renewcommand{\arraystretch}{1.05}
\captionof{table}{Fault-tolerant resources for Shor ECDLP (low-depth-optimal), surface-code conservative scenario.}
\label{tab:phys-low-depth-conservative}
\resizebox{\linewidth}{!}{%
\begin{tabular}{@{}r S[table-format=4.0] S[table-format=2.0] S[table-format=1.0] S[table-format=10.0] S[table-format=9.0] S[table-format=10.0] S[table-format=7.0] S[table-format=7.2] S[table-format=3.2]@{}}
\toprule
\multicolumn{1}{c}{Bits} & \multicolumn{1}{c}{LogicalQ} & \multicolumn{1}{c}{d} & \multicolumn{1}{c}{Factories} & \multicolumn{1}{c}{T} & \multicolumn{1}{c}{T-depth} & \multicolumn{1}{c}{Full depth} & \multicolumn{1}{c}{PhysicalQ} & \multicolumn{1}{c}{Time [s]} & \multicolumn{1}{c}{Time [h]} \\
\midrule
6 & 83 & 13 & 9 & 359272 & 66804 & 344390 & 91366 & 67.16 & 0.02 \\
8 & 102 & 15 & 8 & 699632 & 120598 & 775034 & 127969 & 174.38 & 0.05 \\
12 & 152 & 15 & 8 & 1533528 & 247654 & 1623404 & 163125 & 365.27 & 0.10 \\
16 & 202 & 17 & 8 & 2803768 & 437760 & 2929320 & 254681 & 746.98 & 0.21 \\
24 & 300 & 17 & 7 & 6608608 & 996672 & 7831138 & 334156 & 1996.94 & 0.55 \\
32 & 399 & 19 & 4 & 17507104 & 2485460 & 37667343 & 495247 & 10735.19 & 2.98 \\
48 & 583 & 19 & 5 & 46937844 & 6762858 & 76342091 & 714103 & 21757.50 & 6.04 \\
64 & 760 & 19 & 6 & 102028400 & 14796496 & 139390846 & 925062 & 39726.39 & 11.04 \\
80 & 935 & 21 & 7 & 190387160 & 27672540 & 233945868 & 1385016 & 73692.95 & 20.47 \\
96 & 1111 & 21 & 7 & 321275280 & 46755480 & 368491714 & 1627566 & 116074.89 & 32.24 \\
112 & 1286 & 21 & 8 & 501995340 & 73071460 & 549945995 & 1882519 & 173232.99 & 48.12 \\
128 & 1464 & 21 & 8 & 745509824 & 108618448 & 791826221 & 2127825 & 249425.26 & 69.28 \\
144 & 1645 & 21 & 7 & 1012119456 & 146034429 & 1293177486 & 2363484 & 407350.91 & 113.15 \\
160 & 1822 & 21 & 7 & 1361856276 & 196748605 & 1636092986 & 2607412 & 515369.29 & 143.16 \\
176 & 1997 & 23 & 7 & 1805086480 & 261055620 & 2088627522 & 3417009 & 720576.50 & 200.16 \\
192 & 2174 & 23 & 8 & 2343601220 & 339279894 & 2637488728 & 3726144 & 909933.61 & 252.76 \\
208 & 2351 & 23 & 7 & 2865540864 & 412543714 & 3584950952 & 4002216 & 1236808.08 & 343.56 \\
224 & 2527 & 23 & 7 & 3569725632 & 514684140 & 4296269609 & 4293166 & 1482213.02 & 411.73 \\
240 & 2705 & 23 & 6 & 4379276688 & 626175051 & 6132009701 & 4570891 & 2115543.35 & 587.65 \\
256 & 2884 & 23 & 7 & 5276241436 & 756915745 & 6992537566 & 4883331 & 2412425.46 & 670.12 \\
\bottomrule
\end{tabular}%
}
\end{center}

%% file: resource_estimator/table_phys_lowT_aggressive.tex
\begin{center}
\small
\setlength{\tabcolsep}{5pt}
\renewcommand{\arraystretch}{1.05}
\captionof{table}{Fault-tolerant resources for Shor ECDLP (low-T-optimal), surface-code aggressive scenario.}
\label{tab:phys-low-t-aggressive}
\resizebox{\linewidth}{!}{%
\begin{tabular}{@{}r S[table-format=4.0] S[table-format=2.0] S[table-format=1.0] S[table-format=10.0] S[table-format=9.0] S[table-format=10.0] S[table-format=6.0] S[table-format=5.2] S[table-format=2.2]@{}}
\toprule
\multicolumn{1}{c}{Bits} & \multicolumn{1}{c}{LogicalQ} & \multicolumn{1}{c}{d} & \multicolumn{1}{c}{Factories} & \multicolumn{1}{c}{T} & \multicolumn{1}{c}{T-depth} & \multicolumn{1}{c}{Full depth} & \multicolumn{1}{c}{PhysicalQ} & \multicolumn{1}{c}{Time [s]} & \multicolumn{1}{c}{Time [h]} \\
\midrule
6 & 80 & 7 & 4 & 195488 & 87444 & 302398 & 8624 & 1.86 & 0.00 \\
8 & 93 & 9 & 4 & 366936 & 153478 & 639634 & 15993 & 5.07 & 0.00 \\
12 & 138 & 9 & 4 & 781128 & 312314 & 1310188 & 22008 & 10.38 & 0.00 \\
16 & 184 & 9 & 4 & 1370872 & 533232 & 2265124 & 28156 & 17.94 & 0.00 \\
24 & 275 & 9 & 4 & 3317080 & 1182736 & 6255214 & 40318 & 49.54 & 0.01 \\
32 & 353 & 11 & 4 & 7940148 & 2985143 & 12880141 & 75800 & 124.68 & 0.03 \\
48 & 516 & 11 & 4 & 23410288 & 8707239 & 37577613 & 108343 & 363.75 & 0.10 \\
64 & 679 & 11 & 4 & 51639052 & 19118454 & 82123936 & 140886 & 794.96 & 0.22 \\
80 & 847 & 11 & 3 & 93142380 & 31237410 & 182158912 & 173097 & 1763.30 & 0.49 \\
96 & 1007 & 11 & 4 & 152101800 & 52615020 & 278225126 & 206372 & 2693.22 & 0.75 \\
112 & 1167 & 11 & 4 & 232843492 & 82155766 & 405419577 & 238316 & 3924.46 & 1.09 \\
128 & 1328 & 13 & 4 & 339371488 & 121402416 & 570251987 & 377749 & 6523.68 & 1.81 \\
144 & 1488 & 13 & 4 & 474731964 & 171411804 & 777796744 & 422365 & 8897.99 & 2.47 \\
160 & 1648 & 13 & 4 & 642795480 & 233648140 & 1032088344 & 466981 & 11807.09 & 3.28 \\
176 & 1808 & 13 & 4 & 847289652 & 309493536 & 1339149882 & 511597 & 15319.87 & 4.26 \\
192 & 1968 & 13 & 4 & 1091572176 & 400240728 & 1704197069 & 556213 & 19496.01 & 5.42 \\
208 & 2134 & 13 & 4 & 1373989200 & 481019036 & 2397057497 & 602502 & 27422.34 & 7.62 \\
224 & 2294 & 13 & 4 & 1701999116 & 600748568 & 2909262635 & 647118 & 33281.96 & 9.24 \\
240 & 2454 & 13 & 4 & 2079599368 & 739011267 & 3493715381 & 691734 & 39968.10 & 11.10 \\
256 & 2617 & 13 & 4 & 2507025032 & 858685991 & 4579925011 & 737186 & 52394.34 & 14.55 \\
\bottomrule
\end{tabular}%
}
\end{center}

%% file: resource_estimator/table_phys_lowT_conservative.tex
\begin{center}
\small
\setlength{\tabcolsep}{5pt}
\renewcommand{\arraystretch}{1.05}
\captionof{table}{Fault-tolerant resources for Shor ECDLP (low-T-optimal), surface-code conservative scenario.}
\label{tab:phys-low-t-conservative}
\resizebox{\linewidth}{!}{%
\begin{tabular}{@{}r S[table-format=4.0] S[table-format=2.0] S[table-format=1.0] S[table-format=10.0] S[table-format=9.0] S[table-format=10.0] S[table-format=7.0] S[table-format=7.2] S[table-format=3.2]@{}}
\toprule
\multicolumn{1}{c}{Bits} & \multicolumn{1}{c}{LogicalQ} & \multicolumn{1}{c}{d} & \multicolumn{1}{c}{Factories} & \multicolumn{1}{c}{T} & \multicolumn{1}{c}{T-depth} & \multicolumn{1}{c}{Full depth} & \multicolumn{1}{c}{PhysicalQ} & \multicolumn{1}{c}{Time [s]} & \multicolumn{1}{c}{Time [h]} \\
\midrule
6 & 80 & 13 & 6 & 195488 & 87444 & 302398 & 73938 & 58.97 & 0.02 \\
8 & 93 & 13 & 5 & 366936 & 153478 & 639634 & 75522 & 124.73 & 0.03 \\
12 & 138 & 15 & 5 & 781128 & 312314 & 1310188 & 132188 & 294.79 & 0.08 \\
16 & 184 & 15 & 5 & 1370872 & 533232 & 2265124 & 164531 & 509.65 & 0.14 \\
24 & 275 & 17 & 5 & 3317080 & 1182736 & 6255214 & 293516 & 1595.08 & 0.44 \\
32 & 353 & 17 & 5 & 7940148 & 2985143 & 12880141 & 363959 & 3284.44 & 0.91 \\
48 & 516 & 19 & 5 & 23410288 & 8707239 & 37577613 & 638519 & 10709.62 & 2.97 \\
64 & 679 & 19 & 6 & 51639052 & 19118454 & 82123936 & 833684 & 23405.32 & 6.50 \\
80 & 847 & 19 & 5 & 93142380 & 31237410 & 182158912 & 1011928 & 51915.29 & 14.42 \\
96 & 1007 & 21 & 5 & 152101800 & 52615020 & 278225126 & 1456678 & 87640.91 & 24.34 \\
112 & 1167 & 21 & 5 & 232843492 & 82155766 & 405419577 & 1677178 & 127707.17 & 35.47 \\
128 & 1328 & 21 & 5 & 339371488 & 121402416 & 570251987 & 1899056 & 179629.38 & 49.90 \\
144 & 1488 & 21 & 5 & 474731964 & 171411804 & 777796744 & 2119556 & 245005.97 & 68.06 \\
160 & 1648 & 21 & 5 & 642795480 & 233648140 & 1032088344 & 2340056 & 325107.83 & 90.31 \\
176 & 1808 & 21 & 6 & 847289652 & 309493536 & 1339149882 & 2574338 & 421832.21 & 117.18 \\
192 & 1968 & 21 & 6 & 1091572176 & 400240728 & 1704197069 & 2794838 & 536822.08 & 149.12 \\
208 & 2134 & 21 & 5 & 1373989200 & 481019036 & 2397057497 & 3009825 & 755073.11 & 209.74 \\
224 & 2294 & 23 & 5 & 1701999116 & 600748568 & 2909262635 & 3874925 & 1003695.61 & 278.80 \\
240 & 2454 & 23 & 5 & 2079599368 & 739011267 & 3493715381 & 4139425 & 1205331.81 & 334.81 \\
256 & 2617 & 23 & 5 & 2507025032 & 858685991 & 4579925011 & 4408884 & 1580074.13 & 438.91 \\
\bottomrule
\end{tabular}%
}
\end{center}

%% file: resource_estimator/table_phys_lowW_aggressive.tex
\begin{center}
\small
\setlength{\tabcolsep}{5pt}
\renewcommand{\arraystretch}{1.05}
\captionof{table}{Fault-tolerant resources for Shor ECDLP (low-width-optimal), surface-code aggressive scenario.}
\label{tab:phys-low-width-aggressive}
\resizebox{\linewidth}{!}{%
\begin{tabular}{@{}r S[table-format=4.0] S[table-format=2.0] S[table-format=1.0] S[table-format=12.0] S[table-format=11.0] S[table-format=12.0] S[table-format=6.0] S[table-format=7.2] S[table-format=3.2]@{}}
\toprule
\multicolumn{1}{c}{Bits} & \multicolumn{1}{c}{LogicalQ} & \multicolumn{1}{c}{d} & \multicolumn{1}{c}{Factories} & \multicolumn{1}{c}{T} & \multicolumn{1}{c}{T-depth} & \multicolumn{1}{c}{Full depth} & \multicolumn{1}{c}{PhysicalQ} & \multicolumn{1}{c}{Time [s]} & \multicolumn{1}{c}{Time [h]} \\
\midrule
6 & 73 & 9 & 4 & 1219248 & 681502 & 1985088 & 13320 & 15.72 & 0.00 \\
8 & 81 & 9 & 4 & 2923648 & 1549213 & 4928927 & 14390 & 39.04 & 0.01 \\
12 & 104 & 9 & 4 & 10576320 & 5714012 & 18396215 & 17464 & 145.70 & 0.04 \\
16 & 140 & 11 & 4 & 24381504 & 12862075 & 41735519 & 33275 & 404.00 & 0.11 \\
24 & 204 & 11 & 4 & 91887936 & 49043944 & 162011278 & 46053 & 1568.27 & 0.44 \\
32 & 272 & 11 & 4 & 213387776 & 112164661 & 370730941 & 59629 & 3588.68 & 1.00 \\
48 & 400 & 13 & 4 & 782709504 & 417236048 & 1386923804 & 118976 & 15866.41 & 4.41 \\
64 & 529 & 13 & 4 & 1847084544 & 974006378 & 3245111419 & 154948 & 37124.07 & 10.31 \\
80 & 660 & 13 & 4 & 3730328320 & 1951019013 & 6589083354 & 191477 & 75379.11 & 20.94 \\
96 & 788 & 13 & 4 & 6792749568 & 3618861471 & 12202606456 & 227170 & 139597.82 & 38.78 \\
112 & 916 & 13 & 4 & 10804354624 & 5705909402 & 19274921335 & 262863 & 220505.10 & 61.25 \\
128 & 1045 & 13 & 4 & 16169045504 & 8533547732 & 28814259958 & 298834 & 329635.13 & 91.57 \\
144 & 1173 & 13 & 4 & 23423467968 & 12284917007 & 41544572820 & 334527 & 475269.91 & 132.02 \\
160 & 1301 & 15 & 4 & 32606983040 & 17102545545 & 57765267187 & 492896 & 762501.53 & 211.81 \\
176 & 1429 & 15 & 4 & 44707681920 & 23622083172 & 79757333906 & 540416 & 1052796.81 & 292.44 \\
192 & 1557 & 15 & 4 & 59453297664 & 31693335486 & 107113902705 & 587936 & 1413903.52 & 392.75 \\
208 & 1688 & 15 & 4 & 75674664832 & 40084968281 & 138130068111 & 636570 & 1823316.90 & 506.48 \\
224 & 1816 & 15 & 4 & 94707940096 & 50064207795 & 172786673006 & 684090 & 2280784.08 & 633.55 \\
240 & 1944 & 15 & 4 & 116556830400 & 61530475118 & 212240142701 & 731610 & 2801569.88 & 778.21 \\
256 & 2073 & 15 & 4 & 141721438208 & 74902098856 & 258487044076 & 779501 & 3412028.98 & 947.79 \\
\bottomrule
\end{tabular}%
}
\end{center}

%% file: resource_estimator/table_phys_lowW_conservative.tex
\begin{center}
\small
\setlength{\tabcolsep}{5pt}
\renewcommand{\arraystretch}{1.05}
\captionof{table}{Fault-tolerant resources for Shor ECDLP (low-width-optimal), surface-code conservative scenario.}
\label{tab:phys-low-width-conservative}
\resizebox{\linewidth}{!}{%
\begin{tabular}{@{}r S[table-format=4.0] S[table-format=2.0] S[table-format=1.0] S[table-format=12.0] S[table-format=11.0] S[table-format=12.0] S[table-format=7.0] S[table-format=8.2] S[table-format=5.2]@{}}
\toprule
\multicolumn{1}{c}{Bits} & \multicolumn{1}{c}{LogicalQ} & \multicolumn{1}{c}{d} & \multicolumn{1}{c}{Factories} & \multicolumn{1}{c}{T} & \multicolumn{1}{c}{T-depth} & \multicolumn{1}{c}{Full depth} & \multicolumn{1}{c}{PhysicalQ} & \multicolumn{1}{c}{Time [s]} & \multicolumn{1}{c}{Time [h]} \\
\midrule
6 & 73 & 15 & 5 & 1219248 & 681502 & 1985088 & 86484 & 446.64 & 0.12 \\
8 & 81 & 15 & 5 & 2923648 & 1549213 & 4928927 & 92109 & 1109.01 & 0.31 \\
12 & 104 & 17 & 5 & 10576320 & 5714012 & 18396215 & 139081 & 4691.03 & 1.30 \\
16 & 140 & 17 & 5 & 24381504 & 12862075 & 41735519 & 171594 & 10642.56 & 2.96 \\
24 & 204 & 19 & 5 & 91887936 & 49043944 & 162011278 & 286544 & 46173.21 & 12.83 \\
32 & 272 & 19 & 5 & 213387776 & 112164661 & 370730941 & 363256 & 105658.32 & 29.35 \\
48 & 400 & 21 & 5 & 782709504 & 417236048 & 1386923804 & 620156 & 436881.00 & 121.36 \\
64 & 529 & 21 & 5 & 1847084544 & 974006378 & 3245111419 & 797934 & 1022210.10 & 283.95 \\
80 & 660 & 23 & 5 & 3730328320 & 1951019013 & 6589083354 & 1173719 & 2273233.76 & 631.45 \\
96 & 788 & 23 & 5 & 6792749568 & 3618861471 & 12202606456 & 1385319 & 4209899.23 & 1169.42 \\
112 & 916 & 23 & 5 & 10804354624 & 5705909402 & 19274921335 & 1596919 & 6649847.86 & 1847.18 \\
128 & 1045 & 23 & 5 & 16169045504 & 8533547732 & 28814259958 & 1810172 & 9940919.69 & 2761.37 \\
144 & 1173 & 23 & 5 & 23423467968 & 12284917007 & 41544572820 & 2021772 & 14332877.62 & 3981.35 \\
160 & 1301 & 23 & 5 & 32606983040 & 17102545545 & 57765267187 & 2233372 & 19929017.18 & 5535.84 \\
176 & 1429 & 25 & 5 & 44707681920 & 23622083172 & 79757333906 & 2888672 & 29909000.21 & 8308.06 \\
192 & 1557 & 25 & 5 & 59453297664 & 31693335486 & 107113902705 & 3138672 & 40167713.51 & 11157.70 \\
208 & 1688 & 25 & 5 & 75674664832 & 40084968281 & 138130068111 & 3394531 & 51798775.54 & 14388.55 \\
224 & 1816 & 25 & 5 & 94707940096 & 50064207795 & 172786673006 & 3644531 & 64795002.38 & 17998.61 \\
240 & 1944 & 25 & 5 & 116556830400 & 61530475118 & 212240142701 & 3894531 & 79590053.51 & 22108.35 \\
256 & 2073 & 25 & 5 & 141721438208 & 74902098856 & 258487044076 & 4146484 & 96932641.53 & 26925.73 \\
\bottomrule
\end{tabular}%
}
\end{center}